\documentclass[%
 reprint,
 amsmath,amssymb,
 aps,
]{revtex4-2}

\usepackage[utf8]{inputenc}

\usepackage{graphicx}
\usepackage{dcolumn}
\usepackage{bm}
\usepackage{siunitx}
\usepackage{chemformula}
\usepackage{txfonts}
\usepackage{amsmath}
\usepackage{braket}
\usepackage{float}
\usepackage{color} 
\usepackage[dvipsnames]{xcolor}
\usepackage{silence}
\usepackage{hyperref}

\hypersetup{    
    urlcolor=blue,
}

\begin{document}

\preprint{APS/123-QED}

\title{Coherence Protection for Mobile Spin Qubits in Silicon}

\author{J. A. Krzywda$^{1}$}
\thanks{These authors contributed equally }
\author{Y. Matsumoto$^{2}$}
\thanks{These authors contributed equally }
\author{M. De Smet$^{2}$}
\thanks{These authors contributed equally }
\author{L. Tryputen$^{3}$}
\author{S.L. de Snoo$^{2}$}
\author{S.V. Amitonov$^{3}$}
\author{E. van Nieuwenburg$^{1}$}
\author{G. Scappucci$^{2}$}
\author{L.M.K. Vandersypen$^{2}$}
\email{L.M.K.Vandersypen@tudelft.nl}

\affiliation{$^{1}$$\langle aQa^L
\rangle$ Applied Quantum Algorithms, Lorentz Institute and Leiden Institute of Advanced Computer Science,
Leiden University, The Netherlands \\
$^{2}$QuTech and Kavli Institute of Nanoscience, Delft University of Technology, Lorentzweg 1, 2628 CJ Delft, The Netherlands
\\
$^{3}$QuTech and Netherlands Organization for Applied Scientific Research (TNO), Delft, The Netherlands}

\date{\today}

\begin{abstract}
Mobile spin qubit architectures promise flexible connectivity for efficient quantum error correction and relaxed device layout constraints, but their viability rests on preserving spin coherence during transport. While shuttling transforms spatial disorder into time-dependent noise, its net impact on spin coherence remains an open question. Here we demonstrate systematic noise mitigation during spin shuttling in a linear $^{28}$Si/SiGe quantum dot device. First, by passively reducing magnetic field gradients, we minimize charge-noise coupling to the spin and double the spatially averaged dephasing time $T_2^*(x_n)$ from $4.4$ to $8.5\,\mu\text{s}$. Next, we exploit motional narrowing by periodically shuttling the qubit, achieving a further enhancement in coherence time up to $T_{2}^{*,sh} = 11.5\,\mu\text{s}$. Finally, we incorporate dynamical decoupling techniques while periodically shuttling over distances exceeding $200\,\text{nm}$, reaching $T_\text{2}^{H,sh}= 32\,\mu\text{s}$. For the same setup, we demonstrate that dressed-state shuttling provides robust protection against low-frequency noise with a decay time $T_R^{\text{sh}} = 21\,\mu\text{s}$, without the overhead of pulsed control and allowing protection during one-way spin transport. By preserving coherence over timescales exceeding typical gate and readout operations, the demonstrated strategies establish mobile spin qubits as a viable solution for scalable silicon quantum processors.

\end{abstract}

\maketitle

\section{\label{sec:Introduction}Introduction}
Spin qubits in semiconductor quantum dots are a prominent platform for quantum computation \cite{vandersypen_quantum_2019}, with substantial advancements in recent years. These include demonstrations of high-fidelity single- and two-qubit gate operations \cite{yoneda_quantum-dot_2018,xue_quantum_2022,noiri_fast_2022, Wu_2025_simultaneous, George_2025_12qubit}, extended coherence times \cite{veldhorst_addressable_2014, struck2020low}, and initial progress towards the fabrication of larger qubit arrays \cite{maurand_cmos_2016,zwerver_qubits_2022, wang_operating_2024, George_2025_12qubit, john2025robust, PRX_Ha_HRL}. Despite these achievements, current architectures based on stationary spin qubits face inherent limitations in connectivity and scalability, necessitating the exploration of alternative approaches capable of overcoming these constraints.

Mobile spin qubit architectures offer a distinct paradigm that potentially addresses several fundamental challenges in quantum processor design \cite{taylor_shuttle_2005,vandersypen_interfacing_2017,kunne2024spinbus,Siegel2025snakes, ginzel2024scalable, németh2025omnidirectionalshuttlingavoidvalley}. By enabling the transport of qubits across the device, these architectures present key conceptual advantages over their stationary counterparts. They facilitate flexible qubit connectivity, which is crucial for the efficient implementation of quantum error correction codes often requiring non-local interactions~\cite{bravyi_high-threshold_2024, xu_constant-overhead_2024, siegel2024towards, yenilen2025performance}. Moreover, by physically separating qubit storage regions from designated operation zones, they alleviate constraints on local gate fan-out and increase operational uniformity \cite{matsumoto2025mobile, kunne2024spinbus, patomaki2024pipeline, Ginzel_prb25}.

The performance of quantum processors with mobile spin qubits will highly depend on the fidelity of coherent qubit transport. Intuitively, shuttling is expected to induce errors as the qubit moves through spatially varying charge and magnetic disorder \cite{langrock_blueprint_2023, zhangPRB_2025}. Strong electrostatic disorder along the shuttling path can even cause charge transfer failure\,\cite{krzywda_interplay_2021}.
 
While this risk may be reduced by conveyor-mode shuttling with large voltage amplitudes ~\cite{seidler_conveyor-mode_2022,xue_sisige_2024}, preserving spin coherence remains a challenge. Shuttling at high speed may cause transitions to excited orbital or valley states \cite{langrock_blueprint_2023}, leading to spin relaxation \cite{huang2013spin_relax, huang2014spin_valley, volmer2025reduction, volmer_mapping_2023} and dephasing \cite{pazhedath2025large, langrock_blueprint_2023} through intrinsic spin–orbit coupling or magnetic field gradients. Conversely, slow adiabatic shuttling exposes qubits to dephasing arising from nuclear spins~\cite{assali2011hyperfine} or charge noise in the presence of longitudinal magnetic field gradients~\cite{struck2020low}. This highlights a fundamental trade-off in the speed of spin transport between adiabatic motion and spin dephasing.

Beyond this trade-off, the dephasing process of moving spins is further complicated by spatially correlated noise fluctuations~\cite{yoneda2023noise, rojas2023spatial, rojas2025inferring,  donnelly2024noise}, inducing spatiotemporal correlations tied to their trajectory ~\cite{mokeev2024modeling}. At the same time, the interplay between spatiotemporally correlated noise and shuttling dynamics presents opportunities to exploit specific coherence protection mechanisms. For instance, a qubit encoded in the two-spin singlet and $m=0$ triplet state provides intrinsic protection against correlated noise, so this encoding will enhance the qubit robustness when two electrons are shuttled along the same trajectory rapidly one after another~\cite{zhangPRB_2025}. For single-qubit shuttling, motional narrowing can be induced by moving through a spatially varying environment, effectively averaging out low-frequency fluctuations. This effect was first observed in nuclear magnetic resonance~\cite{hendrickson1973phenomenological} and recently seen in electron spin qubits~\cite{mortemousque_enhanced_2021, struck_spin-epr-pair_2024}. 
Further protection can be obtained by combining motion with active control techniques. This includes shuttling in regimes of high spin-orbit interaction or large magnetic-field gradients~\cite{bosco2024high}, or incorporating dynamical decoupling pulses during the shuttling sequence~\cite{noiri_shuttling-based_2022, yoneda_coherent_2021,van_riggelen-doelman_coherent_2023,desmet_highfidelity2025}.  Despite successful demonstrations of coherent spin shuttling in several semiconductor platforms~\cite{fujita_coherent_2017,flentje_coherent_2017,yoneda_coherent_2021,struck_spin-epr-pair_2024,desmet_highfidelity2025,matsumoto2025mobile, van_riggelen-doelman_coherent_2023}, systematic strategies to mitigate noise during transport remain under-explored.

\begin{figure*}[t!]
\includegraphics[width=1\textwidth]{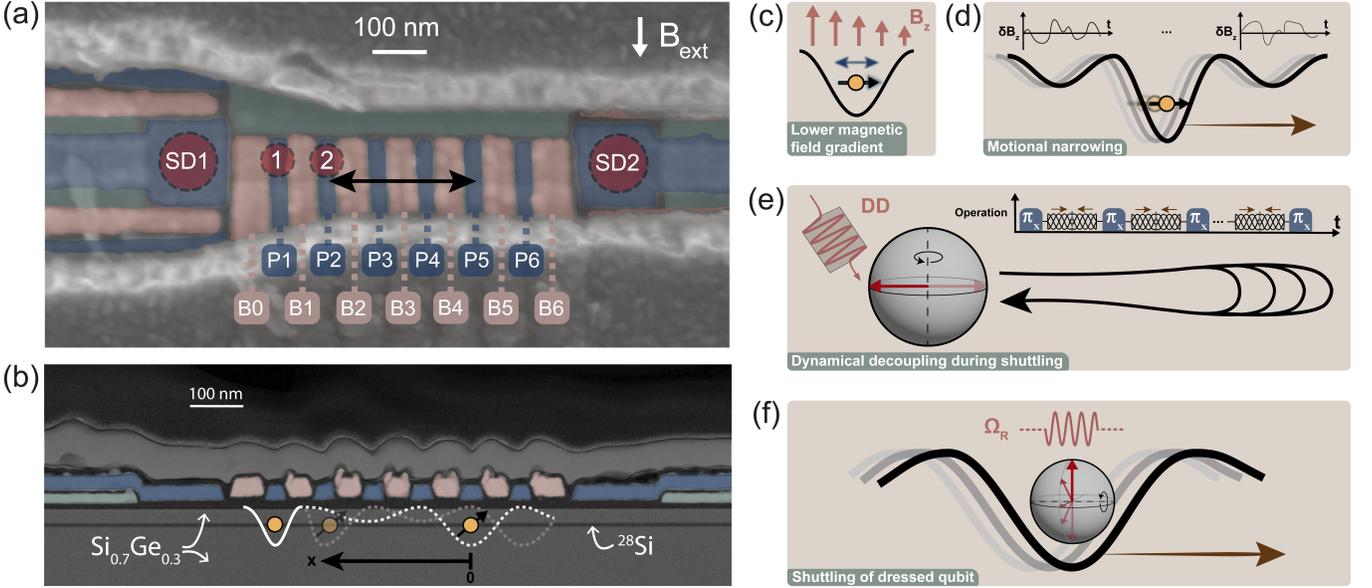}
\caption{ 
Schematic of the experimental setup and protocol using a mobile electron spin as a quantum sensor. 
(a) Top-view false-colored scanning electron microscope image of a Si/SiGe device nominally identical to the one used, showing the linear quantum dot array (static dots 1 and 2 are indicated), plunger gates (P1-P6), barrier gates (B0-B6), and sensing dots (SD1, SD2) and a cobalt micromagnet (gray). An external magnetic field $B_{\text{ext}}$ is indicated. 
(b) Cross-sectional transmission electron microscope image showing the gate stack and $^{28}$Si/SiGe heterostructure. The same false colors are used as in panel (a) for each gate layer, except that here, the gray material on top of the metallization layers is a Pt cap added for improved imaging instead of the micromagnet. A schematic of the potential landscape shows the start and end positions (white dotted lines) of the conveyor transporting qubit 2. Quantum dot 1 is static and contains three electrons, where the one unpaired spin acts as a reference for readout.
(c-f) Conceptual overview of strategies to mitigate spin decoherence during shuttling. 
(c) Passive stabilization: Reducing the longitudinal magnetic field gradient decreases the coupling of charge/electrical noise to the qubit frequency.
(d) Motional narrowing: Rapid shuttling boosts spin coherence by averaging out spatially varying noise $\delta B_z (x,t)$, effectively modifying the noise power spectral density seen by the qubit.
(e) Dynamical decoupling (DD): Periodic shuttling is combined with DD pulses ($\pi_x$) applied during the stationary intervals between transport segments.
(f) Dressed-state shuttling: Continuous driving ($\Omega_R$) protects the qubit against low-frequency noise during transport.}
\label{fig:Fig1} 
\label{fig:1}
\end{figure*}

Here, we present and demonstrate a broad range of approaches to suppressing noise in mobile spin qubits through systematic characterization and advanced control strategies  (Fig.~\ref{fig:1}). We utilize the qubit as a sensor to map the noise landscape and spatiotemporal correlations, providing the basis for several mitigation techniques. We successively study the effect on spin coherence of lowering the longitudinal magnetic field gradient of the micromagnet, motional averaging and dynamical decoupling. Additionally, we introduce dressed-state shuttling, which continuously protects the qubit against low-frequency noise without the overhead of synchronized control pulses. We validate these findings with newly developed theoretical models extending the filter function formalism \cite{biercuk2011dynamical, cywinski2008enhance} and Floquet theory \cite{shevchenko2010landau}, offering a comprehensive framework for coherent mobile qubits.

\begin{figure}[t!]
    \centering
    \includegraphics[width=0.99\linewidth]{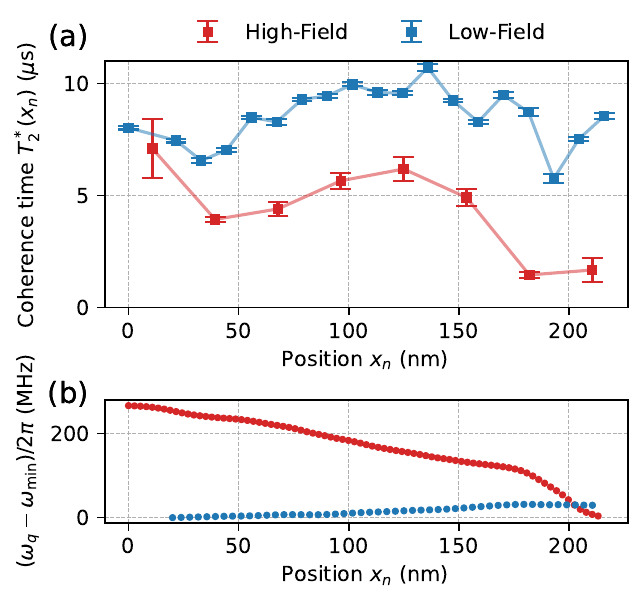}
   \caption{Qubit as a sensor of the noise landscape. 
Results of Ramsey-style dephasing measurements for a stationary qubit as a function of position along the conveyor. 
Red and blue data correspond to the standard \textit{high-field} (\SI{260}{\milli\tesla}) and demagnetized \textit{low-field} (\SI{-30}{\milli\tesla}) setups, respectively. 
(a) Fitted dephasing time, $T_2^*(x_n)$ (solid lines are added for guidance). 
(b) Relative qubit frequency with respect to the corresponding minima $(\omega_q-\omega_\text{min})/2\pi$, along the conveyor obtained through EDSR (high-field) and Ramsey spectroscopy (low-field). For the underlying data, extracted exponents and fits to the decoherence curves, see Appendix~\ref{app:two_points_data}
}
    \label{fig2:noise_features}
\end{figure}

\section{\label{sec:noise_characterization} Noise characterization}
The coherence protection depends on the noise landscape. We first use the qubit as a probe to map the local noise characteristics and then measure spatial correlations across the device (see Fig.~\ref{fig:1} (a) and (b) and the Methods for more details). We thereby distinguish two magnetic field regimes. In the ``high-field'' regime, a magnetic field of 260 mT is applied using a superconducting solenoid. The micromagnet is largely magnetized by a prior sweep of the external magnetic field to 2 T. In the ``low-field'' regime, the magnetic field from the solenoid is swept down and taken to - 30 mT. In this case, the micromagnet is largely demagnetized. 

\subsection{\label{subsec:mapping} Mapping the Noise Landscape}
We begin by using the spin qubit as a static probe of the local noise environment. This is done by preparing a spin superposition state, shuttling it to a specific position $x$ along the conveyor, performing coherent oscillations with a variable waiting time $\tau$, followed by a projection onto the initial state after shuttling the electron back. We assume decoherence is dominated by random fluctuations of wavefunction-averaged spin splitting $\delta \tilde \omega_q(x,t)$, which leads to a random phase accumulated during the waiting time of $\tau$, ${\phi(\tau) = \int_0^\tau \delta \tilde \omega_q(x,t)\text{d}t}$. Any systematic or random phase accumulation from shuttling is $\tau$ independent. By measuring the return probability, we perform a fit to the envelope of the analytical expression,
\begin{equation}
    p_0(\tau) = \frac{A}{2} + \frac{B}{2}\cos(\omega \tau + \phi_0) e^{-\chi(\tau)}, \quad \chi(\tau) = \left(\frac{\tau}{T_2^*}\right)^\alpha,
\end{equation}    
that allows us to extract the central quantity of the paper, the decoherence factor $\chi(\tau)$. It is characterized by two parameters: the coherence decay time $T_2^*(x_n)$ and the decay exponent $\alpha$. As discussed in Appendix~\ref{app:stationary_function}, the latter gives information about the temporal noise correlations, with $\alpha = 1$ indicating white-noise dominated behavior, and $\alpha > 1$ pointing at predominantly low-frequency components, up to $\alpha = 2$, in which case the noise can be treated as quasi-static.

We show the results of these stationary measurements versus position along the conveyor channel in Fig.~\ref{fig2:noise_features}. In the high-field regime (red), we observe a strong spatial dependence of the dephasing time $T_2^*(x_n)$, plotted in Fig.~\ref{fig2:noise_features}~(a). The lowest coherence times occur at large $x_n$, where the magnetic field gradients are also larger, see the spin splitting versus position in Fig.~\ref{fig2:noise_features}~(b). This suggests that the predominant dephasing mechanism is due to charge noise leading to spin splitting fluctuations in the presence of a magnetic field gradient. Consistent with this interpretation, we observe almost twice as long dephasing times in the low-field regime ($8.5~\mu$s versus $4.4~\mu$s averaged over position), where the demagnetized micromagnet produces a strongly reduced magnetic field gradient along the conveyor axis. In Appendix~\ref{app:two_points_data}, we show the extracted decay exponents, indicating a low-frequency origin of noise with $\alpha \approx 2$ in the high-field configuration, and typically faster noise with $\alpha \approx 1.5$ in the low-field. 

\begin{figure}[b!]
    \centering

    \includegraphics[width=\linewidth]{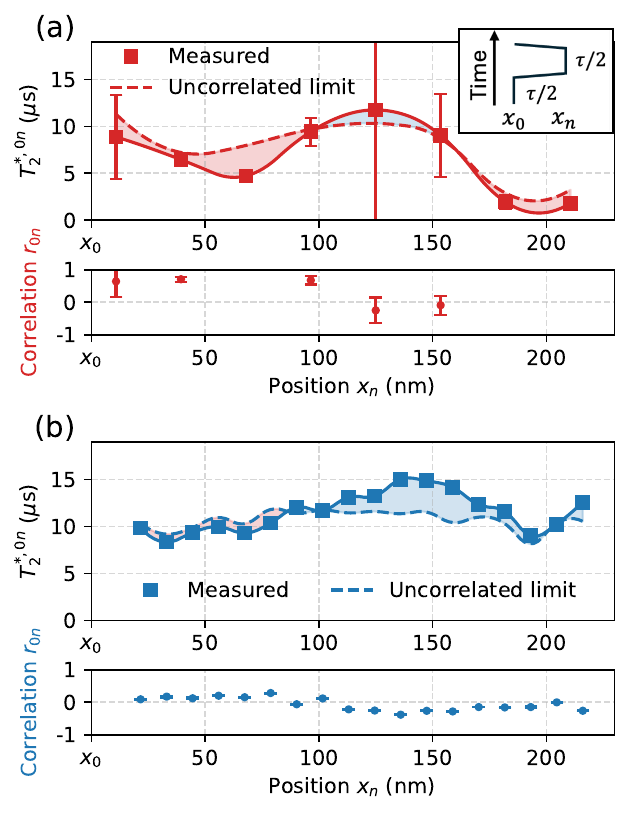}
    \caption{ Spatial noise correlations.
Two-point Ramsey measurement of the dephasing time $T_{2}^{*,0n}(x_n)$ as a function of the variable position $x_n$, with the other point fixed at $x_0$. The fitted dephasing time is plotted as square data points with error bars, along with an interpolated trend (solid line). This is compared to an uncorrelated noise model (dashed line) derived from stationary data. The shaded region is the difference between the two with red color indicating correlations and blue negative correlations. The top/bottom subpanels show the extracted noise correlation coefficient $r_{0n}$. The inset diagram in (a) illustrates the pulse sequence, where an evolution time $\tau/2$ is spent at each location. 
Panel (a) corresponds to the high-field (\SI{260}{\milli\tesla}) case, and panel (b) to the low-field (\SI{-30}{\milli\tesla}) case. For the underlying data and fits to the decoherence curves, see Appendix~\ref{app:two_points_data}.
}
    \label{fig3_twopoint}
\end{figure}

\subsection{\label{subsec:two_point} Spatial noise correlations}
Having mapped the static noise, we next investigate the spatial length scale of the noise correlations, beyond which motional narrowing is activated. We once again perform a Ramsey experiment with the qubit initialized in a superposition state, but split the free evolution time of the qubit equally between two positions: a fixed reference point $x_0$, located on one side of the conveyor, and a variable point $x_n$. We measure the envelope of the return probability and fit the decoherence function $\chi(x_0,x_n,\tau) = (\tau/T_{2}^{*,0n})^{\alpha_{0n}}$ using two free parameters: the two-point decoherence time $T_{2}^{*,0n}$ and the exponent $\alpha_{0n}$. The results are shown in Fig.~\ref{fig3_twopoint}. For comparison, we plot the limit of independent noise using a dashed line, and mark a shaded region indicating positive (red) or negative (blue) correlation. Additionally we use the DC approximation (see Appendix~\ref{app:two_point}):
\begin{equation}  \label{eq:correlated_data}
    \chi_{DC}   (x_0,x_n,\tau) = \left(\frac{\tau/2}{T_{2,0}^*}\right)^{\alpha_0} + \left(\frac{\tau/2}{T_{2,n}^*}\right)^{\alpha_n} + 2r_{0n} \left(\frac{\tau/2}{T_{2,0}^*}\right)^{\alpha_{0}/2} \left(\frac{\tau/2}{T_{2,n}^*}\right)^{\alpha_n/2},
\end{equation}
to fit the correlation coefficients $r_{0n} \in (-1,1)$ using the stationary decay time $T_{2,n}^* \equiv T_2^*(x_n)$ and exponent $\alpha_n \equiv \alpha(x_n)$ at each position $x_n$ (taken from Fig.~\ref{fig2:noise_features}). The resulting $r_{0n}$ are plotted in Fig.~\ref{fig3_twopoint} (b) and (d). The independent noise limit, shown by a dashed line in Fig.~\ref{fig3_twopoint}, is recovered by setting $r_{0n} = 0$.

In the high-field case (Fig.~\ref{fig3_twopoint}~(a)), we report a constant positive correlation $r_{0n} \approx 0.8$ for $x_n < 100$~nm, consistent with previously observed length scales of charge noise correlations in similar devices \cite{yoneda2023noise, rojas2025inferring}. Above $x_n \approx 100$~nm, the two-point signal shows significant beats reflected in a large uncertainty in the fitting procedure, but also is associated with a strong decay of correlation. For large $x_n$ the model \eqref{eq:correlated_data} fails to fit the data, due to the dominant role of high-frequency noise that violates the DC approximation. In Appendix~\ref{app:consistency}, we additionally report that correlations are strongly sensitive to the amplitude of the conveyor pulse. While different conveyor amplitudes result in a similar pattern of stationary noise $T_2^*(x_n)$, variation in both exponents and spatial correlations illustrates a non-trivial interplay between shuttling, the noise landscape and confinement of the quantum dot.

As seen in Fig.~\ref{fig3_twopoint}(b), in the low-field case, the correlation is typically much weaker, with the correlation coefficient staying around $0.1$ for the first $100$~nm. For larger displacements, the correlation becomes significantly negative, until it decays to zero above $170$~nm. The negative correlation could be produced by a local source of electric field fluctuations, which create opposing energy shifts on either side, resulting in anti-correlated noise \cite{rojas2023spatial}. Such a fluctuator would be expected to accelerate dephasing as well, yet no dip in $T_2^*(x_n)$ was observed between the relevant points. Alternatively, the observed correlation pattern might be explained by a rotation of the quantization axis along the channel, which can also turn correlated qubit frequency noise into anti-correlated noise. We theoretically analyze this hypothesis in Appendix~\ref{app:two_point_rotation}, considering both the possibility of diabatic transfer and of fluctuations of the conveyor position in the direction perpendicular to the conveyor motion.

\section{\label{sec:noise_mitigation} Noise mitigation}
In Section~\ref{sec:noise_characterization}, we have shown that qubit dephasing in the conveyor is dominated by fluctuations in the spin splitting with finite spatiotemporal correlations. Here we demonstrate methods to extend coherence times by either breaking or exploiting these correlations.

\subsection{\label{subsec:spatial_mitigation} Breaking Spatial Correlations Through Shuttling}
\begin{figure}[htb!]
    \centering
\includegraphics[width=\linewidth]{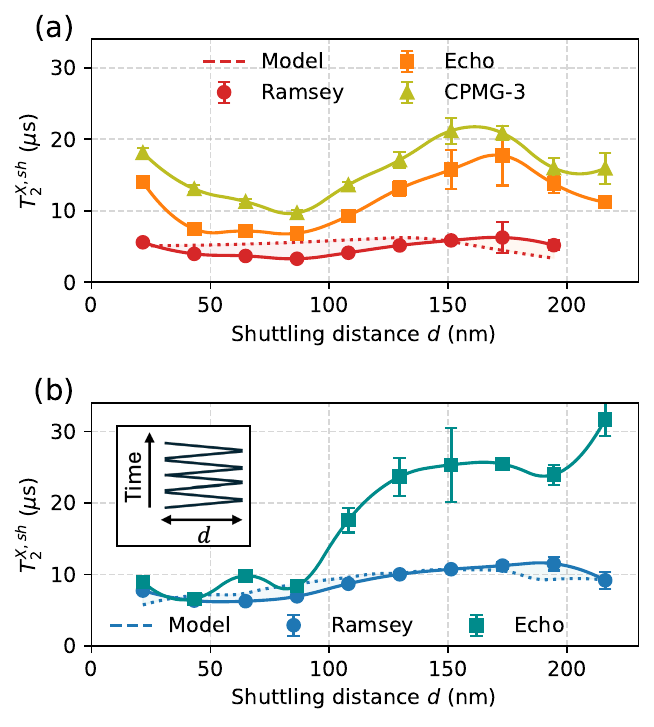}
\caption{Coherence during periodic shuttling. 
The panels show several spin coherence times measured during continuous back-and-forth shuttling, as a function of shuttling distance $d$. The inset in (b) schematically illustrates the shuttling protocol. Measured coherence times are extracted from experiments using a Ramsey-style pulse sequence (circles), a Hahn-echo sequence (squares), and a three-pulse CPMG sequence (triangles), with interpolated lines for guidance. The Ramsey results are compared to a theoretical model (dashed lines) derived from stationary data and a finite correlation length. Panel (a) shows the high-field case, and panel (b) the low-field case. For underlying data and fits to decoherence curves, see Appendix~\ref{app:shuttling_data}.
}
    \label{fig4_shuttling}
\end{figure}

We first investigate the effect of qubit shuttling on the dephasing time. A qubit initialized in a superposition state is now repeatedly shuttled back and forth over a distance $d$ along the conveyor. We measure the return probability as a function of total evolution time $\tau$ (varied by the number of shuttling cycles $N$) and keep the velocity constant at $v=10.8$~m/s. For each shuttling length $d$, we fit the decoherence factor $\chi(d,\tau)$ and extract $T_{2}^{*,sh}(d)$ and $\alpha_\text{sh}(d)$. The results of these experiments are plotted in Fig.~\ref{fig4_shuttling}. We observe that $T_{2}^{*,sh}$ is maximized at large shuttling distances, reaching $\approx 6.2~\mu$s in the high-field case and $\approx 11.5~\mu$s in the low-field case. These values exceed the average stationary dephasing times of $\langle T_2^*(x_n)\rangle \approx 4.4~\mu$s and $8.5~\mu$s reported in Fig.~\ref{fig2:noise_features} for the high-field and low-field regimes respectively. This indicates that shuttling extends coherence time beyond the stationary average case.

In both cases, the improvement can be explained by breaking the spatial correlations achieved through shuttling the electron beyond the correlation length. A similar effect, also referred to as motional narrowing, has been observed in earlier spin shuttling experiments~\cite{mortemousque_enhanced_2021, struck_spin-epr-pair_2024}. However, the extracted exponent $\alpha_{sh}(d) > 1$ (see Appendix~\ref{app:shuttling_data} for the exponent data) indicates the presence of remaining temporal correlations, due to visiting the same region of space multiple times. To understand the underlying mechanism, we connect these shuttling results to the stationary data $T_{2,n}^* \equiv T_{2}^*(x_n)$ and $\alpha_n \equiv \alpha(x_n)$. This analysis uses the coarse-grained model for spatially non-uniform noise developed in Appendix~\ref{app:shuttling_filter}. As shown there, for low-frequency dominated noise, the decoherence factor is given by
\begin{equation}
    \chi^\text{sh}_\text{DC}(d,\tau) = \sum_{n,m=1}^M r_{nm} \left(\frac{\tau/M}{T_{2,n}^*}\right)^{\alpha_n} \left(\frac{\tau/M}{T_{2,m}^*}\right)^{\alpha_m},
    \label{eq:shuttling_decoherence}
\end{equation}
where $M$ is the number of stationary segments within the shuttling range for which $T_2^*(x_n)$ was measured, $r_{nm}$ quantifies the correlation between the segments, and $\tau/M$ represents the time spent in each of them.

In the high-field case (Fig.~\ref{fig4_shuttling}~(a)), we compare the measured $T_{2}^{*,sh}$ (red circles and an interpolated line) with a simple model (dashed line) which assumes uniform spatial correlations given by $r_{nm} = \exp(-|x_n-x_m|/\lambda_c)$ with $\lambda_c = 120$~nm, consistent with the correlation length scale extracted in Fig.~\ref{fig3_twopoint}. This model allows us to qualitatively capture the shape of the experimental $T_{2}^{*,sh}$. In the low-field case (Fig.~\ref{fig4_shuttling}~(b)), the shape of $T_{2}^{*,sh}(d)$ (blue circles and interpolated line) can be closely matched by using a much shorter correlation length $\lambda_c = 15$~nm within the same model (dashed line), which is again consistent with the correlation pattern shown in Fig.~\ref{fig3_twopoint}. This suggests no spatial correlation beyond the size of the electron wavefunction.

As shown in Appendix~\ref{app:shuttling_data}, the measured exponent $\alpha_\text{sh}(d)$ increases with $d$, which indicates a source of effectively high-frequency noise that dominates decoherence for short shuttling distances. This behavior differs from the stationary data (See stationary $\alpha(x_n)$ in Appendix~\ref{app:two_points_data} for comparison) and suggests increased stability in the shuttling path at longer distances $d$. We attribute this stability to the corresponding reduction in shuttling frequency, $\omega_\text{s} = 2\pi v/d$ at fixed $v = 10.8$\,m/s (see discussion in Appendix~\ref{app:two_point_rotation}).

\subsection{\label{subsec:temporal_mitigation} Breaking Temporal Correlation Through Dynamical Decoupling}
Having mitigated spatially correlated noise through shuttling, we now address the residual temporal noise correlations. The shuttling process itself acts as a filter on the noise spectrum; by increasing the shuttling speed, contributions on resonance with the shuttling frequency and its harmonics ($\chi_{AC}$) are pushed to the higher-frequency part of the noise spectra, leaving the quasi-static, or ``DC," component ($\chi_{DC}$) as the dominant dephasing mechanism. This can be understood through the filter function formalism for shuttling, developed in Appendix \ref{app:shuttling_filter}, which decomposes the decoherence function as
\begin{equation}
\chi_{\text{sh}}(d,\tau,\omega_s) = \chi_{DC}^{\text{sh}}(d,\tau) + \chi_{AC}(d,\tau,\omega_s) + \chi_{\text{other}}.
\end{equation}
Standard dynamical decoupling (DD) sequences are specifically designed to counteract low-frequency noise by shaping the control filter to have zero sensitivity at DC (see Appendix~\ref{app:echo_filter} for an illustration). 

To test this, we incorporated echo pulses (a single $\pi$-pulse for Hahn echo and a three-pulse CPMG sequence) applied at the midpoint (or a quarter for CPMG) of the entire shuttling trajectory, immediately before the qubit reverses direction. As shown in Fig.~\ref{fig4_shuttling}, the application of DD yields a two- to threefold improvement in dephasing time, reaching $20~\mu$s in the high-field regime (three-pulse CPMG) and $30~\mu$s in the low-field case (Hahn echo). This dramatic enhancement confirms that low-frequency noise is indeed the primary limiting factor for a qubit in motion once spatial correlations are averaged out by shuttling. 

We observe this improvement for all shuttling distances $d$ in the high-field case, while in the low-field case it occurs only for $d > 100$~nm. For shorter distances, decoupling from low-frequency noise reveals other decoherence mechanisms, $\chi_{\text{other}}$. Based on Fig.~\ref{fig4_shuttling} such mechanisms are strongest around $d = 80$~nm, where the echo improvement is the smallest (high-field) or completely absent (low-field). This behavior could be caused by localized charge defects, resonant coupling, or a rotation of the spin quantization axis during transport. In Appendix~\ref{app:two_point_rotation}, we explore this hypothesis, attributing the short-distance behavior to a combination of high shuttling frequency, magnetic field gradients, and positional instability perpendicular to the motion. We postulate that this same mechanism may drive the locally negative noise correlations observed in Fig.~\ref{fig3_twopoint}.

\begin{figure}[h!]
\includegraphics[width=\columnwidth]{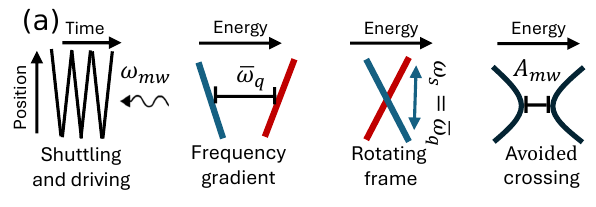}
\vspace{-0.25cm}
\includegraphics[width=\columnwidth]{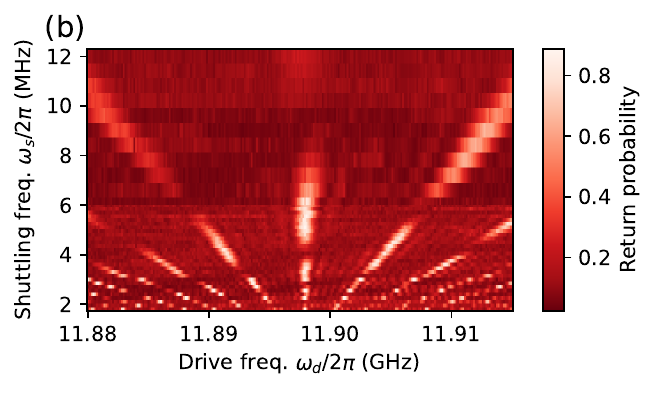}
\vspace{-0.25cm}
\includegraphics[width=\columnwidth]{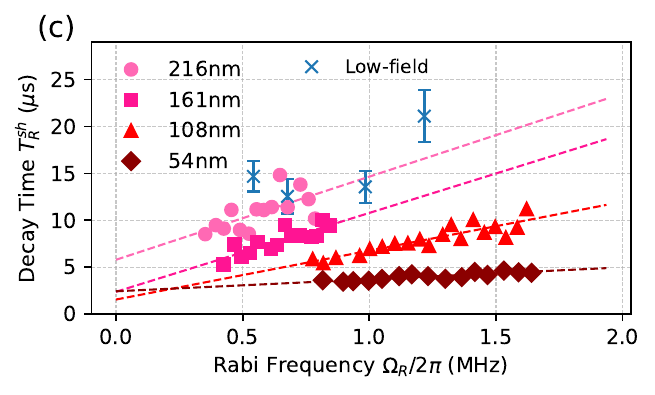}
\vspace{-0.25cm}
\includegraphics[width=\columnwidth]{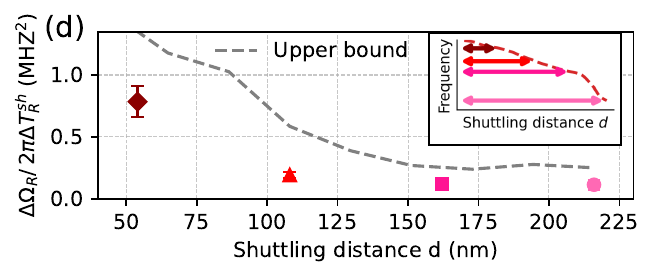}
   
\caption{Dressed-state shuttling experiments. (a) Schematic of the experiment in which we periodically shuttle the electron in the presence of a transverse gradient on resonance with the average spin-splitting. In the rotating frame this results in a periodic drive through an avoided crossing, given by the drive amplitude $A_{mw}$. 
(b) Return probability showing resonant Landau-Zener-Stückelberg-Majorana (LZSM) sidebands as a function of microwave drive and shuttling frequency (high-field).
(c) Measured Rabi decay time $T_R^{sh}$ as a function of the Rabi frequency $\Omega_R/2\pi$ for different shuttling distances in the high-field (red) and low-field regime (blue). In the low-field regime, different $\Omega_R$ are obtained by adjusting the gate voltages.
(d) Ratio $T_R^{sh} / \Omega_R$, extracted from the linear fits in panel (c), as a function of shuttling distance $d$ (points). The dashed line represents a theoretical upper bound derived from $T_2^{CPMG,sh}$ data obtained in the shuttling experiment (see Appendix~\ref{app:coherence_in_driven} for details). The inset relates shuttling distances $d$ to the spin splitting landscape.}
    \label{fig5_dressed}
\end{figure}

\subsection{\label{subsec:dressed} Dressed-State Shuttling for Active Coherence Protection}
While dynamical decoupling is effective for filtering low-frequency noise, it offers protection only at discrete time points and requires precise timing. As an alternative, we explore ``dressed-state shuttling", which provides continuous protection. By driving the qubit during shuttling, we combine the benefits of motional narrowing with the robustness of the dressed basis: any accumulated low-frequency noise is averaged out by the Rabi-like oscillations~\cite{laucht2017dressed, hansen_entangling_2023, tsoukalas2025dressedsinglettripletqubitgermanium}.

In the high-field regime, we implement this concept by applying a microwave drive $H_{mw}(t) = A_{mw} \cos(\omega_d t)\sigma_x$ during shuttling, where the frequency $\omega_d$ is set to the average Larmor frequency over the shuttling trajectory, $ \overline \omega_q$. In the low-field regime, we drive the spin by the resonant motion itself, instead of applying an additional microwave drive \cite{DeSmet_inprep}. As detailed in Appendix~\ref{app:driven_qubit}, in both cases the system is described by the periodic Hamiltonian:
\begin{equation}
\label{eq:avoided_crossing}
H(t) = \frac{\delta  + A_{\parallel} \cos(\omega_s t)}{2} \sigma_z + \frac{A_{\perp}(t)}{2} \sigma_x \,,
\end{equation}
where $\omega_s$ is the shuttling frequency. In the high-field case, under the resonance condition $\delta = \omega_d - \overline{\omega}_q = 0$, we find that $A_\perp(t) = A_{mw}$. The large longitudinal modulation ($A_\parallel \gg A_\perp$) places the system in the Landau-Zener-Stückelberg-Majorana (LZSM) regime \cite{shevchenko2010landau}. The resulting spin response as a function of drive frequency $\omega_d$ (Fig.~\ref{fig5_dressed}~(b)) shows clear sidebands at integer multiples of $\omega_s$, corresponding to constructive interference between periodic passages through the avoided crossing of Eq.~\ref{eq:avoided_crossing}. Conversely, in the low-field case, the drive is intrinsic ($A_\perp(t) \propto \cos(\omega_s t)$) and at resonance ($\omega_s = \overline \omega_q$), it results in standard Rabi oscillations without LZSM interference. We note that although a triangular pulse was implemented in the experiments, we use the cosine model here for simplicity. It contains the main harmonic and trajectory smoothing at the turning points reduces the contribution of higher harmonics.

The performance of the protection against dephasing by continuous driving is summarized in Fig.~\ref{fig5_dressed}~(c), where we plot the Rabi decay time $T_R^{sh}$ against the effective drive strength $\Omega_R$. By adjusting the shuttling velocity to maintain resonance, we achieve decay times up to $T_R^{sh} = 15~\mu$s (high-field) and $T_R^{sh} = 20~\mu$s (low-field). These values are comparable to those achieved with Hahn echoes but are obtained without the overhead of pulsed control sequences.

\begin{figure*}[htb!]
    \centering
    \includegraphics[width=0.99\linewidth]{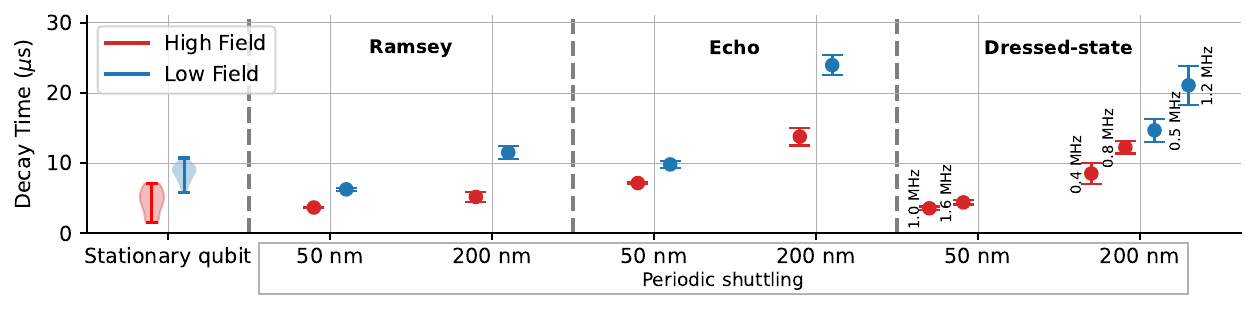}
    \caption{Overview of coherence time improvements. This plot summarizes the gains in coherence achieved through complementary mitigation strategies. We begin with a baseline stationary $T_2^*(x_n) \sim 4.4~\mu$s averaged over a spatial range of 216 nm in the high-field regime (red, leftmost violin plot). Passive noise mitigation by reducing the magnetic field gradient (low-field, blue) immediately yields a factor of two improvement. Periodic shuttling further enhances coherence via motional narrowing, with larger gains observed at increased shuttling distances (150 nm vs 50 nm). Incorporating a Hahn echo pulse during transport effectively filters residual low-frequency noise, extending the spin coherence time up to $\sim 25~\mu$s at $200$nm and $32~\mu$s at $220$nm (not shown here). Finally, we present dressed-state shuttling as a scalable alternative; here, the Rabi decay time $T_R^{sh}$ appears to be roughly proportional to the driving frequency (annotated in MHz) for the various pairs of data points tested, achieving coherence times comparable to those from echo-protected shuttling.}
    \label{fig:app_overview}
\end{figure*}

This enhancement can be understood by transforming to the dressed qubit frame. The drive opens an energy gap $\Omega_R$ along a new quantization axis, effectively converting longitudinal noise into transverse noise. Averaging over one shuttling period yields the effective Floquet Hamiltonian:
\begin{equation}
    H_\text{eff} = \frac{\Omega_\text{R}}{2} \sigma_x + \frac{\delta + \xi_T}{2} \sigma_z,
\end{equation}
where $\xi_T$ represents the noise contribution averaged over the trajectory and $\delta = \omega_s-\overline \omega_q$. Note that this Hamiltonian can be cast into the standard dressed-state formalism by a basis rotation.
As demonstrated in Appendix~\ref{app:driven_qubit}, the effective Hamiltonian represents effective dynamics at multiples of the period, which at high field averages out the evolution away from the avoided crossing, while at low field, it neglects the doubly rotating term.

At resonance ($\delta = 0$), we derive the effective Rabi decay time in Appendix~\ref{app:coherence_in_driven} as
\begin{equation}
    T_R^{sh} \approx  \frac{2\Omega_\text{R}}{\langle \xi_T^2 \rangle}.
\end{equation}
We validate this model in Fig.~\ref{fig5_dressed}~(d) by plotting the quantity $\Delta \Omega_R/\Delta T_R^{sh}$. The data shows qualitative agreement with the effective formula:
\begin{equation}
    \label{eq:Tr_tshuttle}
    \frac{\Delta \Omega_R}{\Delta T_R^{sh}} = \frac{\langle \xi_T^2\rangle}{2} \leq \frac{v}{2d\, T_2^{CPMG,sh}}.
\end{equation}
This relation, derived in Appendix~\ref{app:coherence_in_driven}, assumes that high-frequency noise dominates decoherence over the short duration of a single period, $T_s = 2d/v \ll T_2^{*,sh}(x_n)$. Consequently, we treat the decay rate measured in the DD-protected experiment as an upper bound on the high-frequency noise, i.e. ${1/T_{\text{fast}}^{sh} \leq 1/T_{2}^{CPMG,sh}}$. This allows us to derive an upper bound for the noise variance $\langle \xi_T^2\rangle = 2\chi(T_s)/T_s^2 \leq 2(T_s/T_2^{CPMG,sh})/T_s^2$  depicted as a dashed line in Fig.~\ref{fig5_dressed}~(d).

While Eq.~\ref{eq:Tr_tshuttle} suggests a positive relation between distance and decay time for a fixed Rabi frequency ($T_R^{sh} \propto \Omega_R \,d$), increasing the shuttling distance can actually reduce $T_R^{sh}$ by suppressing $\Omega_R$. To quantify this trade-off, we note that for a harmonic drive at the primary resonance ($k=0$), the Rabi frequency scales as:
\begin{equation}
    \Omega_\text{R} \approx A_{mw} J_0\left(\frac{d^2 \overline{\nabla B_\parallel}}{2\pi v} \right),
\end{equation}
where $J_0$ is the zeroth-order Bessel function and $\overline{\nabla B_\parallel}$ is the average gradient. As seen in the high-field data and analyzed further in Appendix~\ref{app:coherence_in_driven}, $\Omega_R$ decreases and eventually oscillates around zero as $d$ increases. This highlights the importance of minimizing longitudinal gradients: their absence would prevent the suppression of $\Omega_R$ (keeping $J_0 \approx 1$), allowing the system to operate in a regime of optimal protection with $\Omega_R \sim A_\text{mw}$.

\section{\label{sec:Discussion}Discussion and Summary}
Our results, summarized in Fig.~\ref{fig:app_overview}, provide a systematic framework for mitigating decoherence in mobile spin qubit architectures. We have demonstrated three complementary strategies for noise mitigation, which when applied together, enable the preservation of coherent quantum information for tens of microseconds.

First, passive mitigation via micromagnet demagnetization boosted the baseline (stationary) $T_2^*(x_n)$ by approximately a factor of two, through reduction of the system's sensitivity to charge noise. Second, we demonstrated that transforming a stationary qubit into a moving one improves coherence, evidenced by an additional factor of $\sim 1.5$ improvement in the position-averaged dephasing time. We attribute this improvement to the motional narrowing effect, which acts as a spatiotemporal filter that averages out noise once the shuttling distance exceeds the noise correlation length. We observed this enhancement for shuttling distances beyond $\sim100$~nm, a length scale consistent with previous reports of charge fluctuations in Si/SiGe heterostructures. 

Finally, we successfully demonstrated a mobile qubit that preserves a coherent state for $>20\,\mu$s. To achieve this, we incorporated active driving to filter low-frequency noise associated with periodic traversals of the same location. We realized two approaches, dynamical decoupling pulses at symmetric points and continuous spin driving. With a single echo pulse at low-field, we achieved a $32\,\mu$s coherence time when shuttling over 225 nm. For the dressed-state approach we measured a Rabi oscillation decay time of $15\,\mu$s at high-field for a similar shuttling distance, and of $20\,\mu$s at low-field. The Rabi decay is a measure of how well the phase of a dressed qubit is preserved, and can be compared with the Ramsey or echo decay for qubits in the standard basis. A rigorous quantitative comparison of how well arbitrary qubit states in the standard and dressed basis are preserved, would require quantum process tomography or an equivalent procedure. Nevertheless, coherence times of tens of microseconds significantly relax fault-tolerance timing constraints, providing the necessary margin for using mobile qubits in high-weight parity checks, and extending the capabilities recently demonstrated in~\cite{Undseth2026weight}.

Of these two active methods, the dressed-state approach offers a more flexible alternative, protecting the qubit continuously without requiring precisely timed pulses. We note also that the application of an external microwave drive (or utilizing a transverse magnetic field gradient) can effectively mimic the effect of the much stronger spin-orbit interaction in Ge, which has also been proposed for noise mitigation via quantization axis rotations during shuttling \cite{bosco2024high}.

Consistent with theoretical models, we find a proportionality between the Rabi decay time and the Rabi frequency $T_R^{sh} \propto \Omega_R$, and highlighted a key trade-off associated with increasing shuttling distance. While longer distances extend coherence through motional averaging, the benefits are limited by longitudinal gradients that eventually lower the effective Rabi frequency $\Omega_R$. In particular, we showed that when gradients exceed the size of the avoided crossing (defined by the driving amplitude), the spin state of the mobile qubit undergoes LZSM interference. This opens the possibility for further optimization regarding the interplay between driving frequency, shuttling speed, and gradient strength, as described by the Bessel functions.

As we have shown, these different techniques can be deployed individually or in combination to match specific operational scenarios. Minimizing longitudinal magnetic field gradients along the conveyor is universally beneficial for charge-noise-dominated systems, though it presents trade-offs for individual qubit addressability. Noise landscape mapping and pulsed DD sequences like CPMG are ideal for closed-loop transport, such as in ``quantum bus" operations where a qubit is moved to a specific location and back. In contrast, dressed-state shuttling is much more flexible and can be used in scenarios where precise pulse timing might be difficult. Furthermore, this method can be extended to open-loop, unidirectional transport over long distances. In such a case, one could utilize multiple driving fields or exploit natural transverse gradients. Together, these methods provide a practical toolkit for preserving quantum information during shuttling, offering a clear direction for integrating these protocols into large-scale device architectures.

In summary, we have demonstrated powerful strategies to mitigate decoherence for a mobile spin qubit, boosting its coherence time by nearly an order of magnitude. This was achieved by systematically characterizing the spatiotemporal noise environment and then applying targeted mitigation techniques. The characterization revealed a charge noise correlation length of approximately \SI{100}{\nano\meter} with spectral characteristics ranging from white to $1/f$ noise. Guided by these insights, we enhanced spin coherence from a baseline ${T_2^*(x_n) \approx \SI{4.4}{\micro\second}}$ to a maximum of \SI{32}{\micro\second} using motional averaging combined with dynamical decoupling. Furthermore, we showed that dressed-state shuttling offers a robust alternative, yielding a decay time of \SI{20}{\micro\second}. Together, these methods provide a practical toolkit for preserving quantum information during shuttling, which will accelerate the development of scalable and robust spin-based quantum processors.

\section{Methods}

\subsection{Experimental Procedures}

Experiments were performed on a device fabricated on an isotopically purified $^{28}$Si/SiGe heterostructure with integrated cobalt micromagnets, operated in a dilution refrigerator at a base temperature of $\sim20$ mK.

The experiments are performed on a $^{28}$Si/SiGe heterostructure potentially hosting a linear array of six quantum dots, as shown in Fig.\ref{fig:Fig1}~(a). A cobalt micromagnet is patterned on top of the gate stack, providing a spatially varying stray magnetic field that enables electric-dipole spin resonance (EDSR) for single-qubit control\,\cite{obata_coherent_2010} and produces distinct spin resonance frequencies for electrons in different dots. The Si/SiGe heterostructure, illustrated in Fig.\ref{fig:Fig1}~(b), consists of an undoped Si quantum well between SiGe barriers, and an epitaxial Si cap layer, providing a low-disorder environment for high-fidelity spin operations.

For the measurements reported here, one electron serves as a reference, while a second electron is initialized in dot 2 and subsequently transported through the array. The remaining quantum dots are kept empty in order to isolate and characterize the noise processes acting on the mobile spin. 

\subsubsection{Parity Readout and Initialization Sequence}
We use Pauli spin blockade for readout, employing the (3,1)–(4,0) charge transition for the dot pair (Q1,Q2). We first post-select for odd parity in order to initialize the two qubits in the (1,0) spin state.

\subsubsection{Two-Tone Conveyor Pulse}
To shuttle electrons, we apply shaped voltage pulses to the plunger gates. Following previous work \cite{desmet_highfidelity2025}, we use a two-tone sinusoidal pulse for each gate, $V(t) = V^{DC}_n + \frac{A}{2}\big[\sin(2\pi ft - \phi_n) + \sin(\pi ft - \theta_n)\big]$, which creates robust potential barriers and minimizes charge leakage during transport.

\section*{\label{sec:ackn}Acknowledgments}
We wish to thank S.G.J. Philips for writing control libraries and designing the PCB, R. Schouten, R. Vermeulen, O. Benningshof and T. Orton for support with the measurement setup and dilution refrigerator, and other members of the Vandersypen, Veldhorst, Scappucci, and Dobrovitski groups for fruitful discussions.
We acknowledge financial support from the Army Research Office (ARO) under grant number W911NF2310110. The views and conclusions contained in this document are those of the authors and should not be interpreted as representing the official policies, either expressed or implied, of the ARO or the US Government. The US Government is authorized to reproduce and distribute reprints for government purposes notwithstanding any copyright notation herein. 

\section*{\label{sec:contrib}Author contributions}
Y.M. and M.D.S. performed the experiments and data analysis. J.A.K. developed the theoretical models. Simulations were carried out by J.A.K. Libraries for experimental control were written by S.L.S. and Y.M. Y.M., M.D.S., and L.M.K.V. contributed to data interpretation. L.T. fabricated the device, while S.V.A. refined the device design. A.S. and G.S designed and grew the heterostructure. J.A.K, Y.M., M.D.S., and L.M.K.V. wrote the manuscript with comments by all authors. Y.M conceived the project. L.M.K.V. supervised the project.

\section*{\label{sec:interests}Competing interests}
The authors declare no competing interests.

\section*{\label{sec:contrib}Data and code availability}
The raw measurement data supporting the findings of this work are available in the Zenodo repository at \href{https://doi.org/10.5281/zenodo.18470200}{https://doi.org/10.5281/zenodo.18470200}. The data analysis pipeline and code used to generate the figures can be found on GitHub at \href{https://github.com/jaq-lab/mobile-qubit-protection}{https://github.com/jaq-lab/mobile-qubit-protection}.

\bibliographystyle{naturemag}
\bibliography{manualbib}

\clearpage
\onecolumngrid
\appendix

\newpage

\newcommand{\sinc}{\operatorname{sinc}}

\section{Stationary experiment as a probe of noise spectrum \label{app:stationary_function}
}
In this Appendix, we review the relationship between the noise power spectral density (PSD) and the coherence decay exponent. We adopt the standard filter-function formalism~\cite{cywinski2008enhance, biercuk2011dynamical, Barnes_2016} to provide a theoretical basis for the stationary measurements presented in the main text. 

In a stationary experiment, we move the electron to position $x_n$ and measure the coherence as a function of time $C(\tau)$ using a Ramsey sequence, i.e., we initialize the spin superposition, shuttle to position $x_n$, wait for a time $\tau$, shuttle back and project it into the initial state. The coherence decay is given by $C(\tau) = e^{-\chi(\tau)}$, where the decoherence function $\chi(\tau)$ is half the variance of the accumulated phase, $\chi(\tau) = \frac{1}{2}\langle \phi(\tau)^2 \rangle$. The phase is given as an integral of the noise at a given position:
\begin{equation}
    \phi(\tau) = \int_0^\tau \text{d}t \int_{-\infty}^{\infty} \rho(x-x_n) \delta\omega(x, t) \text{d}x = \int_0^\tau dt' \, \delta\tilde \omega(x_n, t'), 
\end{equation}
where $\delta\tilde \omega(x_n, t) = \int_{-\infty}^{\infty} dx \, \rho(x-x_n) \delta\omega(x, t)$ is the noise averaged over the electron wavefunction $\rho(x)$. 

With those definitions we proceed with a standard treatment of qubit decoherence, starting with expressing the decoherence function in terms of the correlation function i.e.
\begin{equation}
    \chi(x_n,\tau) = \frac{1}{2} \int_0^\tau dt_1 \int_0^\tau dt_2 C(x_n, t_1-t_2),
\end{equation}
which we assume is stationary, i.e. $C(x_n, |t_1-t_2|) = \langle \delta\tilde \omega(x_n, t_1) \delta\tilde \omega(x_n, t_2) \rangle$. This allows us to express the correlation function via its Fourier transform, the Power Spectral Density (PSD), denoted $S(x_n,\omega)$ and defined as $C(x_n, t) = \int_{-\infty}^{\infty} \frac{d\omega}{2\pi} S(x_n, \omega) e^{i\omega t}$, which gives the standard filter function expression:
\begin{equation}
    \chi(x_n,\tau) = \frac{1}{2} \int_{-\infty}^{\infty} \frac{d\omega}{2\pi} S(x_n, \omega) |F(\omega, \tau)|^2,
\end{equation}
where $F(\omega, \tau) = \int_0^\tau dt \,e^{-i\omega t} = \tau e^{-i\omega \tau/2} \sinc\left(\frac{\omega \tau}{2}\right)$ is the filter function for a Ramsey sequence. Substituting the filter function and assuming an even PSD, the expression becomes:
\begin{equation}
    \chi(x_n,\tau) = \frac{\tau^2}{2\pi} \int_0^\infty d\omega \, S(x_n, \omega) \sinc^2\left(\frac{\omega \tau}{2}\right).
\end{equation}
We concentrate on common noise spectra that follow a power-law dependence $S(x_n, \omega) = A_n/\omega^{\beta_n}$, where $A_n$ is the noise amplitude at position $x_n$ and $\beta_n$ is the noise exponent. This form captures a wide range of physical noise processes, including white noise ($\beta_n = 0$) and $1/f$-noise ($\beta_n = 1$). 

For the cases when $\beta_n < 1$, the integral is convergent in the low-frequency limit, and can be directly computed as:
\begin{equation}
    \chi(x_n,\tau; \beta < 1) = \frac{A_n \tau^{1+\beta_n}}{2\pi} I_\beta, \quad I_\beta = \int_0^\infty du \, \frac{1}{u^{\beta_n}} \sinc^2\left(\frac{u}{2}\right).
\end{equation}
where $I_\beta$ is a dimensionless constant that depends only on $\beta_n$. The coherence   $C(\tau)$ at position $x_n$ thus falls off with an exponent $\alpha_n = 1+\beta_n$.

However for $\beta_n > 1$ the integral diverges at low frequencies, indicating that the noise is dominated by very low-frequency components. In such a case the integral can be evaluated by assuming the low-frequency cutoff $\omega_\text{min} \ll 1/\tau$ associated with the total time of experiment. This leads to the approximate expression

\begin{equation}
    \chi(x_n,\tau;\beta > 1) \approx \frac{A_n \tau^2}{2\pi} \int_{\omega_\text{min}}^{1/\tau} d\omega \, \frac{1}{\omega^\beta}
\end{equation}
For $\beta > 1$, the integral evaluates to $\frac{1}{\beta-1} \left( (\omega_\text{min})^{1-\beta} - \tau^{\beta-1} \right)$. In the quasi-static limit ($\omega_\text{min} \ll 1/\tau$), the $(\omega_\text{min})^{1-\beta}$ term dominates, making the integral a constant independent of $\tau$. This results in $\chi(\tau) \propto \tau^2$, and thus a coherence decay with exponent $\alpha = 2$.

For $\beta_n=1$ the integral diverges logarithmically in both limits, which introduces a weak logarithmic correction to the quadratic decay \cite{cywinski2008enhance}. However, this correction is often negligible in practical scenarios, and the decay can still be approximated as $\chi(\tau) \propto \tau^2$. 

Combining these regimes, we recover the relation maping the measured decay exponent $\alpha_n$ to the underlying noise exponent $\beta_n$, $\alpha_n \approx \min(2, \, 1 + \beta_n)$. 
This implies that measuring the coherence decay form:
\begin{equation}
    \chi(x_n, \tau) = \begin{cases}
    (\tau/T_2^*)^{\beta_n + 1}, & \beta_n < 1 \\
    (\tau/T_2^*)^2 & \beta_n \geq 1 \\
    \end{cases}
\end{equation}
allows us to directly distinguish between white noise ($\alpha_n=1$) and $1/f$-type noise ($\alpha_n \approx 2$). We note that while this mapping assumes a single dominant power-law noise source, experimentally observed exponents in the range $1 < \alpha_n < 2$ can also result from a competition between white noise (e.g. thermal or readout) and low-frequency $1/f$ noise, or from the logarithmic correction associated with $\beta_n=1$ noise over finite timescales.

\newpage

\section{Two-Point Correlation in the Quasi-Static Limit \label{app:two_point}}
In this Appendix, we derive the decoherence function for a split Ramsey experiment performed across two spatial locations. This derivation generalizes the standard single-qubit filter function formalism~\cite{cywinski2008enhance} to include spatial cross-correlations, which in spin qubits were detected in decoherence of singlet and triplet two-qubit state \cite{boter2020spatial} and recent  spectroscopy experiments ~\cite{yoneda2023noise, rojas2023spatial}. However in contrast to these works, instead of using many qubits we use a single one that probes different position in space. This modification introduces a short delay, which in principle mixes spatial and temporal correlations. 

The protocol consists of evolving the qubit at position $x_n$ for a duration $\tau/2$, shuttling it to position $x_m$ in a time $\Delta t$, and then evolving at $x_m$ for another duration $\tau/2$. The accumulated random phase $\delta \phi$ is:
\begin{equation}
    \delta \phi = \underbrace{\int_0^{\tau/2} dt' \, \delta\tilde\omega(x_n, t')}_{\phi_n} + \underbrace{\int_{\tau/2 + \Delta t}^{\tau + \Delta t} dt' \, \delta\tilde\omega(x_m, t')}_{\phi_m}
\end{equation}
We neglect decoherence during the fixed shuttling time $\Delta t$. The decoherence function $\chi = \frac{1}{2}\langle \delta \phi^2 \rangle = \frac{1}{2}\langle (\delta \phi_n + \delta \phi_m)^2 \rangle$ is given by:
\begin{equation}
    \chi(x_n, x_m, \tau) = \underbrace{\frac{1}{2} \left\langle\delta  \phi_n^2 \right\rangle}_{\chi(x_n, \tau/2)} + \underbrace{\frac{1}{2} \left\langle \delta \phi_m^2 \right\rangle}_{\chi(x_m, \tau/2)} + \underbrace{\left\langle \delta \phi_n \delta  \phi_m \right\rangle}_{\text{Cross-term}} \;.
\end{equation}
The first two terms are the decoherence functions for waiting for a time $\tau/2$ at each location individually, which relate directly to the spectral density $\chi(x_n,\tau/2) = (\tau/2)^2 \int_{-\infty}^\infty (\text{d}\omega/2\pi) S(x_n, \omega) \sinc^2(\omega\tau/4)$ and hence can be related to the corresponding $T_2^*$ and $\alpha$ parameters through $\chi(x_n, \tau/2) = (\tau/2T_2^*(x)_n))^{\alpha_n}$. The third term is the cross-term, which captures the correlation between the    noise at the two positions.

\subsection{Cross-term evaluation}
The cross-term can be expressed using the spatiotemporal cross-spectral density, $S(x_n, x_m, \omega)$, defined as the Fourier transform of the cross-correlation function $C(x_n, x_m, t) = \langle \delta\tilde\omega_n(t_1) \delta\tilde\omega_m(t_2) \rangle$ with $t = t_1 - t_2$. With this definition, the cross-term is given by:
\begin{equation}
    \left\langle \delta \phi_n \delta  \phi_m \right\rangle  = \int_{-\infty}^{\infty} \frac{d\omega}{2\pi} S(x_n, x_m, \omega) \text{Re} \left[ F_n(\omega) F_m^*(\omega) \right]
\end{equation}
where $F_n(\omega)$ and $F_m(\omega)$ are the filter functions for the evolution at position $x_n$ and $x_m$, respectively. For the time intervals $t \in [0, \tau/2]$ and $t \in [\tau/2 + \Delta t, \tau + \Delta t]$, these are:
\begin{align}
    F_n(\omega) = \int_0^{\tau/2} e^{-i\omega t} dt = e^{-i\omega\tau/4} \frac{\tau}{2} \sinc\left(\frac{\omega\tau}{4}\right), \quad 
    F_m(\omega) = \int_{\tau/2 + \Delta t}^{\tau + \Delta t} e^{-i\omega t} dt = e^{-i\omega(\frac{3\tau}{4} + \Delta t)} \frac{\tau}{2} \sinc\left(\frac{\omega\tau}{4}\right) \;.
\end{align}
The interference term in the filter function is therefore given by:
\begin{equation}
    \text{Re} \left[ F_n(\omega) F_m^*(\omega) \right] = \left| \frac{\tau}{2} \sinc\left(\frac{\omega\tau}{4}\right) \right|^2 \cos\left(\omega\left(\frac{\tau}{2} + \Delta t\right)\right)
\end{equation}
which explicitly shows how the qubit motion can modulate the cross-spectral density, depending on the time delay $\Delta t$ and evolution time $\tau$. 

We now make an additional assumption: that the cross-spectral density can be expressed as the geometric mean of the individual spectral densities, weighted by a normalized correlation coefficient $r_{nm}$:
\begin{equation}
\label{eq:cross_spectrum}
    S(x_n, x_m, \omega) = r_{nm} \sqrt{S(x_n, \omega) S(x_m, \omega)}, \quad |r_{nm}| \leq 1
\end{equation}
For $1/\omega^\beta$ spectra, we have $\sqrt{S(x_n, \omega) S(x_m, \omega)} = \sqrt{A_n A_m}/\omega^{(\beta_n + \beta_m)/2}$, which allows the integral to be performed with the $\sinc^2$ filter. 

\subsection{DC Approximation}
Finally, motivated by the dominant role of low-frequency noise, we consider the quasi-static limit, and compute the contribution to decoherence from the fluctuations that are slow compared to the total evolution time $\tau + \Delta t$. In this case, we can drop the oscillatory cosine term from the cross-term (i.e., $\cos(\dots) \approx 1$). Assuming the noise at both positions has a $1/\omega^\beta$ spectrum, the cross-term integral factorizes, and we compute the low-frequency (DC) contribution to decoherence as:
\begin{align}
\label{eq:chi2_dc}
    \chi_{DC}(x_n, x_m, \tau) &= \chi(x_n, \tau/2) + \chi(x_m, \tau/2) + 2r_{nm,\phi} \sqrt{\chi(x_n, \tau/2) \chi(x_m, \tau/2)} \nonumber \\
    &= \left(\frac{\tau/2}{T_{2,n}^*}\right)^{\alpha_n} + \left(\frac{\tau/2}{T_{2,m}^*}\right)^{\alpha_m} + 2 r_{nm,\phi} \left(\frac{\tau/2}{T_{2,n}^*}\right)^{\alpha_n/2} \left(\frac{\tau/2}{T_{2,m}^*}\right)^{\alpha_m/2},
\end{align}
where $r_{nm,\phi} = \langle \phi_n \phi_m \rangle/ \sqrt{\langle\phi_n^2\rangle \langle \phi_m^2\rangle} $ is the correlation coefficient between the random phases which is related to, but not identical, to the noise correlation $r_{nm}$ from Eq.~\eqref{eq:cross_spectrum}, as we will prove numerically below. In the main text $r_{nm,\phi}$ is fitted to experimental data, and called $r_{nm}$ for brevity. This expression has a simple interpretation: the total decoherence is the sum of the individual contributions from each position, plus a cross-term that depends on the correlation coefficient $r_{nm,\phi} \approx r_{nm}$. In the strict DC limit ($\omega \tau \ll 1$), the accumulated phase is directly proportional to the instantaneous noise amplitude, $\phi(\tau) \approx \tau \delta\tilde\omega$. Consequently, the phase correlation coefficient $r_{nm,\phi}$ becomes analytically identical to the noise correlation $r_{nm}$ defined in Eq.~\eqref{eq:cross_spectrum}.

\subsection{Numerical validation} 
We now perform a numerical test of the DC approximation in Eq. \eqref{eq:chi2_dc}. For modeling purposes, we consider two sources of random noise $\xi_1(t)$ and $\xi_2(t)$, which are contributing to the noise at position $A$ and $B$, i.e.
\begin{equation}
    \delta\tilde\omega_A(t) = v_{1A} \xi_1(t) + v_{2A} \xi_2(t), \quad \delta\tilde\omega_B(t) = v_{1B}\xi_1(t) + v_{2B}\xi_2(t)
\end{equation}
where $v_{ij}$ are coupling coefficients. We generate the noise using an inverse Fourier transform  method \cite{timmer1995generating} from the spectral density $S(\omega) = 1/\omega^{\beta_{i}}$, where $\beta_i \in (0,1)$.

\begin{figure}[htb!]
    \centering
    \includegraphics[width=\linewidth]{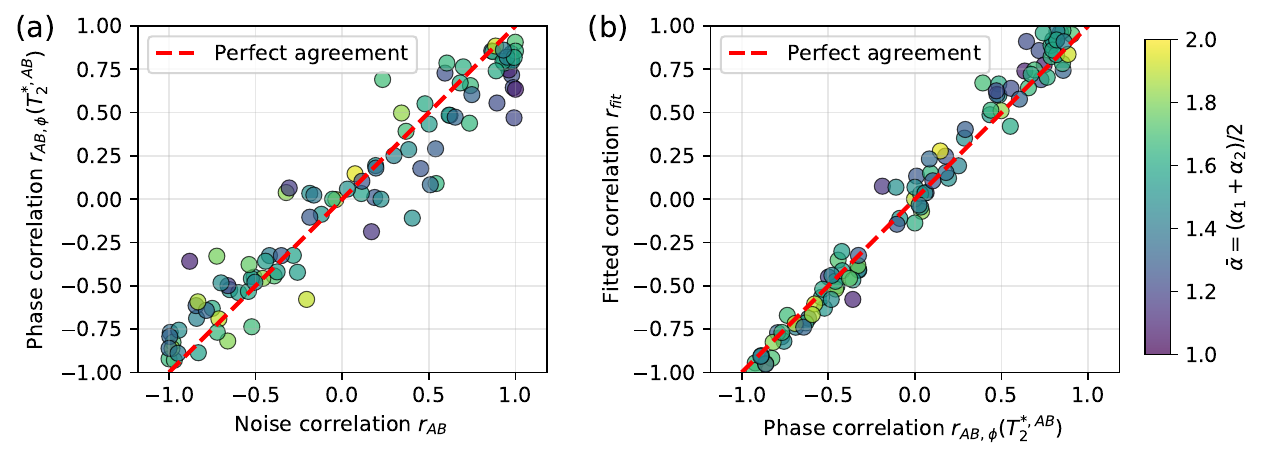}
    \caption{Validation of DC approximation for the two-point case (a) Comparison between the true noise correlation $r_{AB}$, and the phase correlation $r_{AB,\phi}$. (b) Comparison between the fitted correlation coefficient $r_{fit}$ from 1000 randomly drawn realizations of the noise process and the phase correlation $r_{AB,\phi}$. The dots represents independent realizations of source spectra of the form $S_i(\omega) = a_i/\omega^\beta_i $ and couplings $v_{i\alpha}$. The color stands for the average decay exponent from stationary measurements. }
    \label{fig:placeholder}
\end{figure}

To validate the DC approximation, we draw 1000 random realizations of the noise processes and compute the accumulated phases $\delta \phi_A$ and $\delta \phi_B$ for a range of evolution times $\tau$. From these, we numerically calculate the full decoherence $\chi_{\text{total}}(\tau) = \frac{1}{2}\langle (\delta \phi_A + \delta \phi_B)^2 \rangle$ and its components: $\chi_A(\tau) = \frac{1}{2}\langle \delta \phi_A^2 \rangle$, $\chi_B(\tau) = \frac{1}{2}\langle \delta \phi_B^2 \rangle$, and the cross-term $\langle \delta \phi_A \delta \phi_B \rangle$. We fit the total simulated decoherence $\chi_{\text{total}}(\tau)$ and extract a fitted correlation parameter, $r_{\text{fit}}$. Second, we compare against two correlation coefficients 
\begin{equation}
r_\text{AB} = \frac{\langle \delta\tilde\omega_A(t) \delta\tilde\omega_B(t) \rangle}{\sqrt{\langle \delta\tilde\omega_A(t)^2 \rangle \langle \delta\tilde\omega_B(t)^2 \rangle}} \quad \text{and} \quad r_{AB,\phi} = \frac{\langle \delta \phi_A \delta \phi_B \rangle}{\sqrt{\langle \delta \phi_A^2 \rangle \langle \delta \phi_B^2 \rangle}},
\end{equation}
where $r_\text{AB}$ quantifies the correlation of the underlying noise processes, while $r_{AB,\phi}$ quantifies the correlation of the accumulated phases. We evaluate $r_{AB,\phi}$ at $\tau = T_2^{*,AB}$. 

As shown in Fig.~\ref{fig:placeholder}, we find that $r_{\text{fit}} \approx r_{AB,\phi} \approx r_{AB}$. This confirms that in the two-point Ramsey experiment, the parameter extracted from the decoherence fit provides a faithful measure of the spatial correlations of the underlying noise fields.

\newpage

\section{Rotation of quantization axis during two-point transfer \label{app:two_point_rotation}}
In this Appendix, we analyze two distinct physical mechanisms that can lead to the effective anti-correlated phase noise observed in the experiments. These mechanisms represent two different regimes of the magnetic field landscape:

\begin{enumerate}
    \item Large field rotation: The first mechanism relies on a significant rotation of the quantization axis (direction of $|\mathbf{B}|$ vector) between the two shuttling locations. If the magnetic field vector rotates by a sufficiently large angle (approaching $\pi$), while the spin evolution remain diabatic, positive noise correlations in the laboratory frame can project onto opposite signs in the qubit frame.
    
    \item Perpendicular fluctuations: The second mechanism relies on the spatial profile of the magnetic field magnitude $|\mathbf{B}|$. Even if the quantization axis remains (almost) constant, a transverse displacement noise $\delta y$ can generate anti-correlated frequency shifts if the transverse gradient $\partial \omega / \partial y$ changes sign between the two locations. Additionally the movement in the direction perpendicular  the shuttling axis can explain emergence of high-frequency noise at small shuttling distance, where the shuttling frequency is the largest.

\end{enumerate}
\begin{figure}[htb!]
    \centering
    \includegraphics[width=0.9\linewidth]{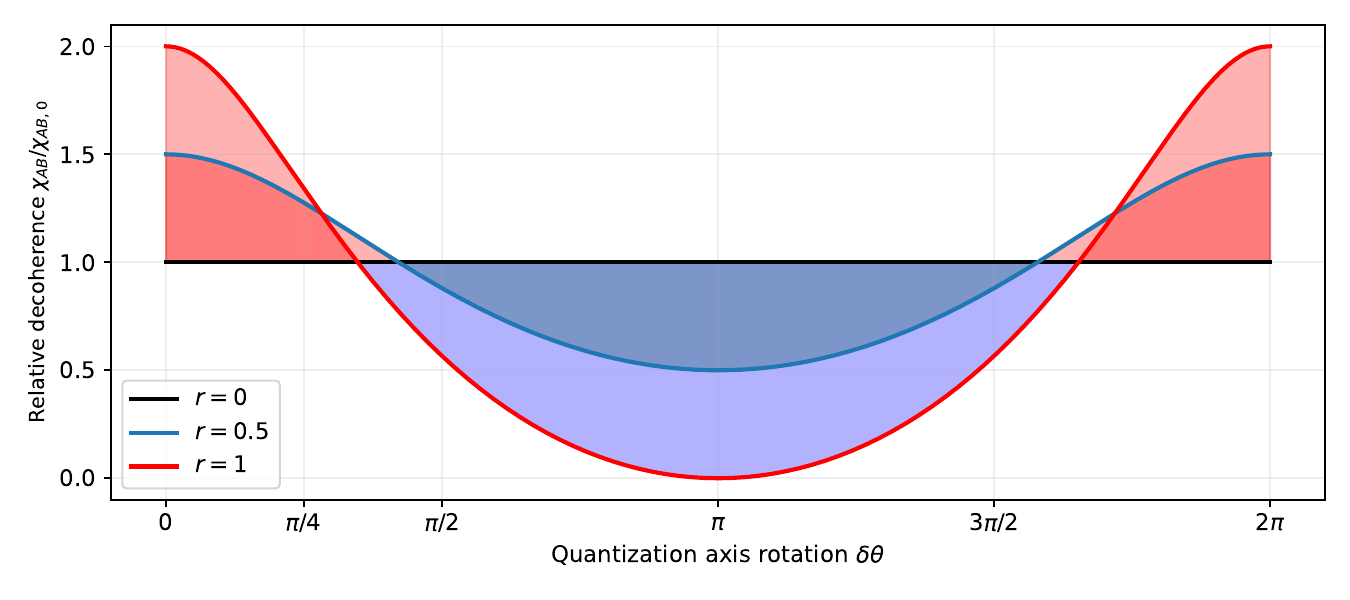}
    \caption{Effect of quantization axis rotation during transfer on two-point Ramsey coherence, in the units of $\chi_{AB,0} \equiv \chi_{AB}(r=0,\tau = T_2^{*,AB})$. Decay of the two-point Ramsey signal as a function of the rotation angle of the quantization axis during the transfer, for uncorrelated noise (black) and correlated noise with $r=0.5$ (blue) and $r =1$ (red). For the last one $\delta \theta\approx 3\pi/8$ is sufficient to turn correlated noise into effectively anti-correlated. Here we set $\chi_A = \chi_B = 0.5$ to compare at the $T_2^{*,AB}$ timescale for uncorrelated case.}
    \label{fig:two_point_rotation}
\end{figure}
In the following sections, we derive the theoretical conditions for both cases and discuss their physical plausibility within the experimental setup.

\subsection{Full rotation of quantization axis combined with diabatic evolution}
Firstly we consider the case where the quantization axis of the noise rotates as a function of position $x$. The noise Hamiltonian in the lab frame is:
\begin{equation}
    H_L(x, t) = [\omega + \delta \omega(x, t)] \left( \cos \theta(x) \frac{\sigma_z}{2} + \sin \theta(x) \frac{\sigma_x}{2} \right) =\big[\omega + \delta \omega(x, t)\big] R_y[\theta(t)] \frac{\sigma_z}{2} R_y^\dagger[\theta(t)]
\end{equation}
We model the two-point experiment where the qubit is held at position $x_1$ for time $\tau/2$, then shuttled to position $x_2$ and held there for time $\tau/2$. For simplicity of the argument, we assume the noise is quasistatic $\delta \omega(x,t) = \delta \omega(x)$ and the transfer is completely diabatic, i.e. the qubit state remains constant during shuttling, which takes place if $\dot \theta(t) \gg \omega$ \cite{langrock_blueprint_2023}. In such a case the evolution operator can be written as:
\begin{equation}
    U_{\text{total}} = U_B(\tau/2) U_A(\tau/2) = e^{-i [\omega+ \delta \omega_B][\cos\theta_B\sigma_z + \sin\theta_B\sigma_x]\tau/2} e^{-i [\omega+ \delta \omega_A][\cos\theta_A\sigma_z + \sin\theta_A\sigma_x]\tau/2}.
\end{equation}
We can now associate the $z$-axis with the quantization axis at position $A$ and express evolution at position $B$ as 
\begin{equation}
R_y^\dagger(\delta \theta) e^{-i [\omega + \delta \omega_B] \sigma_z \tau/2} R_y(\delta \theta),
\end{equation}
where $\delta \theta = \theta(x_B) - \theta(x_A)$ is the relative rotation angle. We assume the initial state is $|+\rangle$ and compute decay of the Ramsey signal. One can neglect the last rotation around the $y$ axis, as it does not contribute to the dephasing. Before this rotation the state is given by:
\begin{equation}
    |\psi\rangle = e^{-i [\omega + \delta \omega_B] \sigma_z \tau/2} R_y(\delta \theta) e^{-i [\omega + \delta \omega_A] \sigma_z \tau/2} |+\rangle = \frac{1}{\sqrt{2}} \begin{pmatrix}
e^{-i(2\omega + \delta\omega_A + \delta\omega_B)\tau/2}\cos(\delta\theta/2) - e^{i(\delta\omega_A - \delta\omega_B)\tau/2}\sin(\delta\theta/2) \\
e^{-i(\delta\omega_A - \delta\omega_B)\tau/2}\sin(\delta\theta/2) + e^{i(2\omega + \delta\omega_A + \delta\omega_B)\tau/2}\cos(\delta\theta/2).
\end{pmatrix}
\end{equation}

In the Ramsey experiment we effectively measure $\langle \sigma_x \rangle$:
\begin{equation}
    S_x = \langle \psi | \sigma_x | \psi \rangle = \cos^2(\delta\theta/2) \cos[(\delta\omega_A + \delta\omega_B)\tau] + \sin^2(\delta\theta/2) \cos[(\delta\omega_A - \delta\omega_B)\tau],
\end{equation}
which after averaging over Gaussian noise $\delta \omega_i$, i.e. $\cos(\delta \omega_A \pm \delta \omega_B) \to e^{-\big(\langle \delta \omega_A^2\rangle^2 +\langle \delta \omega_B^2\rangle^2 \pm 2 r \langle \delta \omega_A^2\rangle\langle \delta \omega_B^2\rangle\big)\tau^2/2}$, and associating $\chi_i = \langle \delta \omega_i^2\rangle \tau^2/2$ gives
\begin{equation}
    \Gamma(\tau, \delta\theta) = \langle S_x \rangle = e^{-(\chi_A + \chi_B + 2 r \sqrt{\chi_A \chi_B})} \cos^2(\delta\theta/2) + e^{-(\chi_A + \chi_B - 2 r \sqrt{\chi_A \chi_B})} \sin^2(\delta\theta/2) \;.
\end{equation}

This clearly shows how the rotation of the quantization axis during the transfer modulates the effect of spatial correlations on the decoherence, and can turn the positively correlated noise into effectively anti-correlated noise. However one has to highlight that in this model effective correlation coefficient depends on both the rotation angle $\delta \theta$ and waiting time. For instance in the limit of small dephasing $\chi_i \ll 1$ one can expand the exponentials to get:
\begin{equation}
    \Gamma(\tau, \delta\theta) \approx 1 - [\chi_A + \chi_B + 2 r \sqrt{\chi_A \chi_B} \cos(\delta\theta)] \;,
\end{equation}
which shows that the effective correlation coefficient is $r_{\text{eff}} = r \cos(\delta \theta)$. Therefore, in the limit of small dephasing, a $\pi/2$ rotation is sufficient to completely suppress the effect of correlations. However this is no longer true if we consider stronger dephasing, typically used for extracting $T_2$. In that case, the effective correlation coefficient depends non-linearly on $\chi_i$. To illustrate the relation between $\delta \theta$ and the noise correlation we concentrate at the effective correlation at the $T_2^{AB}$ timescale for uncorrelated noise, i.e. we set $\chi_A = \chi_B = 0.5$. We compare the uncorrelated case $r=0$ with correlated noise $r>0$ as a function of $\delta \theta$ in Fig.~(\ref{fig:two_point_rotation}). For strong correlation ($r=1$), the effective anti-correlation is visible at $\delta \theta >3\pi/8$. This shows that anti-correlated noise can emerge from positively correlated noise even for moderate rotation angles of the quantization axis. 

We now discuss the physical plausibility of this mechanism in the experimental setup. The main requirement is a substantial rotation of the quantization axis between the two shuttling locations. Given the geometry of the micromagnet and the expected magnetic field profile, achieving a rotation angle of $\delta \theta \gtrsim 3\pi/8$ over a distance of approximately 200 nm appears challenging. Furthermore, while such rotation can explain the anti-correlated noise observed in the two-point Ramsey experiment, it does not naturally account for the behavior seen in the periodic shuttling experiment, where coherence times do not improve with echo pulses at small displacements. This suggests that while quantization axis rotation may contribute to the observed phenomena, it is unlikely to be the sole mechanism at play.

\subsection{Random displacement along y-direction \label{app:two_point_rotation_b}}
\begin{figure}[htb!]
    \centering
    \includegraphics[width=0.98\linewidth]{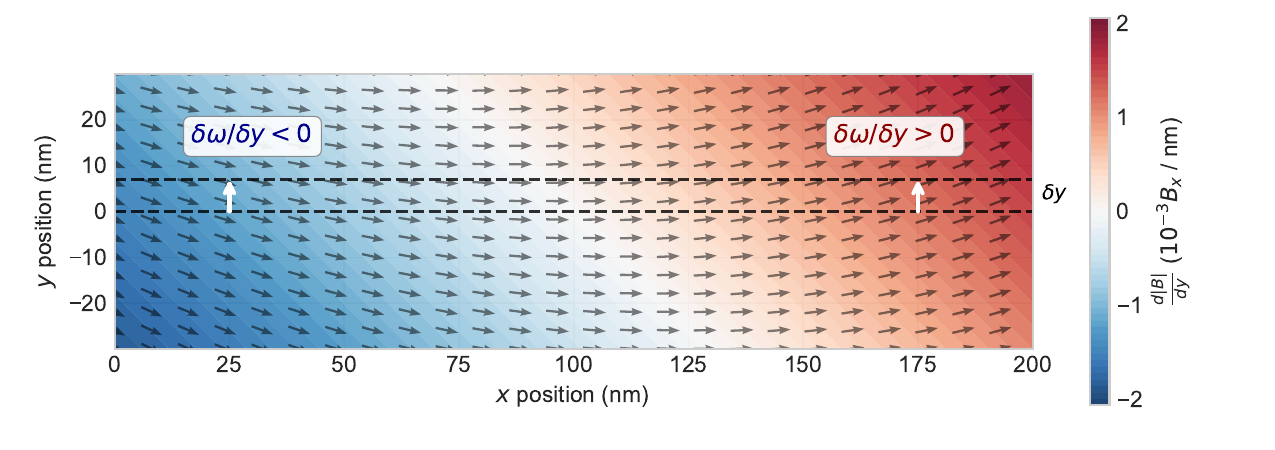}
    \caption{Example of a magnetic field landscape. A visualization of a possible magnetic field pattern along the conveyor that results in anti-correlated noise due to common position fluctuations along $y$. The arrows indicate the local magnetic field direction, while the color map represents the transverse gradient magnitude.}
    \label{fig:magnetic_field_dir}
\end{figure}
While the quantization axis rotation discussed above provides a geometric origin for anti-correlations, it requires substantial rotation angles ($\delta\theta \gtrsim 3\pi/8$) that may exceed those present in the device. Therefore, we investigate an alternative hypothesis where the quantization axis rotations are small, but the qubit frequency is sensitive to random spatial displacements along a direction perpendicular to the main axis of motion.

We model this as a quasistatic shift $\delta y$ along the $y$ direction (e.g., due to imperfect potential control). Assuming the local effective magnetic field has the shape illustrated in Fig.~\ref{fig:magnetic_field_dir}, i.e., $\mathbf{B}(x) = (B_x, B_y(x,y_0))$ where $y_0$ is the nominal position along $y$, the quantization axis is rotated by an angle $\theta(x,y) = \arctan[B_y/B_x]$. Crucially, we focus on the modification to the spin-splitting, given by:
\begin{equation}
    \omega(x,y) = \gamma \sqrt{B_x^2 + [B_y(x, y_0 + \delta y)]^2},
\end{equation}
where $\gamma$ is the gyromagnetic ratio. For $B_x \gg B_y$, this can be expanded to the lowest order in $\delta y$ as:
\begin{equation} \omega(x,y) \approx \gamma B_x + \frac{\gamma [B_y(x, y_0)]^2}{2 B_x} + \frac{\gamma B_y(x-x_0, y_0)}{B_x} \frac{\partial B_y}{\partial y} \delta y \approx \omega_0(x,y_0) + f(x)\,\delta y
\end{equation}

This shows that the random displacement along $y$ introduces an additional noise term proportional to $\delta y$ with a position-dependent prefactor $f(x) = \gamma \tfrac{B_y}{B_x} \tfrac{\partial B_y(x,y_0)}{\partial y}$. Crucially, if the gradient $\tfrac{\partial B_y(x,y_0)}{\partial y}$ changes sign between the two positions $x_A$ and $x_B$, the additional noise term will be effectively anti-correlated. 

We now relate this to the two-point Ramsey experiment described in the main text and Fig.~\ref{fig3_twopoint}. One explanation for the observed pattern is a dominant contribution from $\delta y$ displacements, combined with a sensitivity function $f(x)$ that changes continuously from a positive value for $x < 100$~nm, crossing zero near $x \approx 100$~nm, to a negative value beyond that point. Indeed, the position $x \approx 100$~nm coincides with a region of relatively high $T_2^*(x_n)$, consistent with measured in the single-point Ramsey experiment, consistent with a first-order sweet spot where the sensitivity vanishes. Furthermore, this sign change explains the transition from positive correlations (below 100~nm) to negative correlations (above 100~nm). This hypothesis is also consistent with the arc-like spatial dependence of $T_2^*$ observed in the stationary qubit experiment (Fig.~\ref{fig2:noise_features}).

This model only partially explains the periodic shuttling experiment, which reports effectively uncorrelated noise for small displacements ($d < 100$ nm). In particular, those displacements are not large enough to bridge the regions of opposite sign in $f(x)$ to provide decoupling via sign-cancellation. However, during periodic shuttling, the displacement likely becomes a time-dependent function $\delta y(t)$. If $\delta y(t)$ varies rapidly enough compared to the shuttling period, effectively undergoing motional narrowing of the transverse jitter, its contribution to the accumulated phase averages out:
\begin{equation}
    \delta \phi = \int_0^\tau f(x(t)) \delta y(t) dt \approx 0.
\end{equation}

In such a case, decoherence would be dominated by other noise sources or by high-frequency components of the $\delta y$ noise, explaining the lack of coherence time improvement after Hahn-echo pulses. For small displacements, the shuttling frequency $\omega_s = 2\pi v/d$  is higher, possibly contributing to rapid fluctuations in electron position and inducing inelastic orbital and valley transitions. We hypothesize that for larger displacements ($d > 100$ nm), the qubit trajectory is more stable, leading to a more quasistatic $\delta y$ during the Ramsey time, which could explain the observed behavior. Eventually, we expect that side of the conveyor to be characterized by either more stable magnetic field gradients or sources of decoherence other than $\delta y$ noise, leading to the observed enhancement of coherence time with echo pulses at large displacements.
\newpage
\section{Periodic Shuttling as a Filter Function: Coarse-Grained model}
\label{app:shuttling_filter}

In this section we develop a general framework to model decoherence during periodic shuttling through a spatially varying noise landscape. The problem of dephasing during shuttling was already considered in \cite{langrock_blueprint_2023}, and very recently in \cite{bosco2024high,mokeev2024modeling,zhangPRB_2025}. While \cite{bosco2024high} have implemented the filter function formalism for spatially uncorrelated noise, \cite{mokeev2024modeling,zhangPRB_2025} have explicitly studied the interplay between spatial and temporal correlations affecting the qubit during shuttling. The latter works, however, concentrated on the theoretical model where spatial correlations are homogeneous and independent from the temporal noise of finite correlation time. Here, motivated by the experimental findings we combine the filter function formalism with spatially inhomogeneous noise, i.e. where the noise correlation in space is not only a function of a distance $K(x,x')\neq K(|x-x'|)$. In contrast to previous works, we avoid using correlation functions which are not well defined for a power-law spectra $S(\omega) = 1/\omega^\beta$, and construct a coarse-grained model that allows us to capture the effect of spatially varying noise characteristics on the decoherence during periodic shuttling.

\begin{figure}[htb!]
    \includegraphics[width=0.99\linewidth]{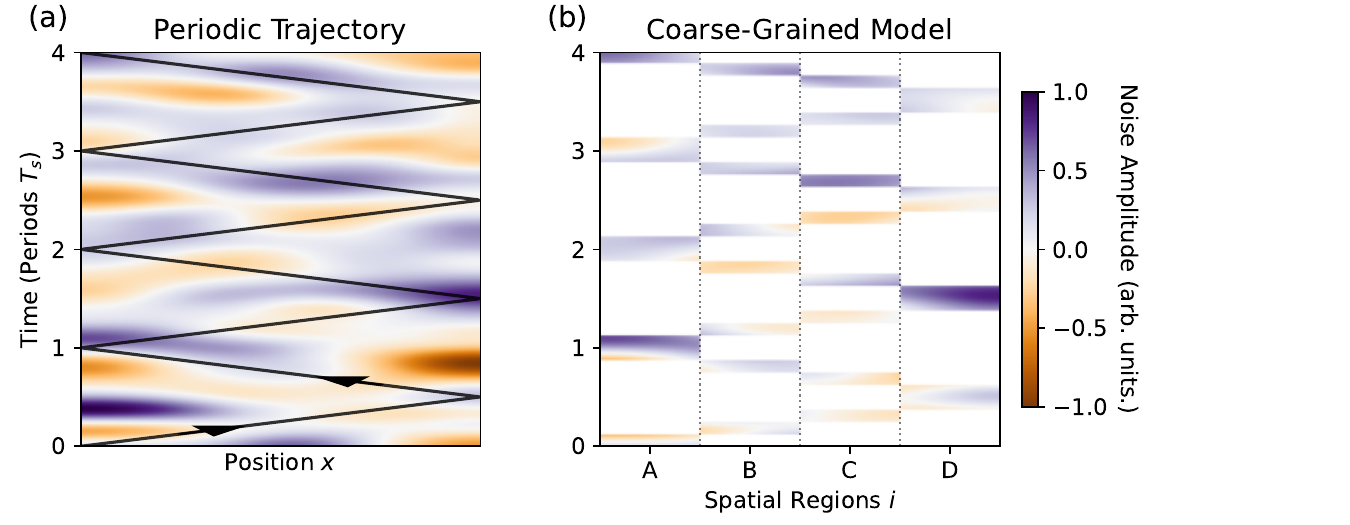}
    \caption{Illustration of the coarse-graining procedure for a spatially inhomogeneous noise landscape. (a) A realization of continuous fluctuations in spin-splitting, where the spatial correlation length varies with position. The solid black line traces the periodic trajectory of the qubit over three periods. (b) The corresponding coarse-grained model with  spatial regions (labeled A, B, C, D). The effective noise is constructed by sampling the continuous landscape only when the trajectory resides in a specific region, effectively mapping the continuous motion onto the discrete sequence of noise operators  used to calculate the structure factors. For illustration purposes, we created an artificial noise pattern by applying a Gaussian filter to uncorrelated noise in space and time, leading to the creation of correlation patches.}
    \label{fig:coarse_grained}
\end{figure}

We start by the analysis of periodic shuttling. For a pure dephasing model, the phase accumulated during periodic shuttling is given by:
\begin{equation}
    \delta \phi(\tau) = \int_0^{\tau} \text{d}t' f(t') \int \text{d}x  \,\rho\big(x- x(t')\big) \,\delta\omega(x, t') ,
\end{equation}
where $\tau$ is the total evolution time, $x(t) = x(t+T_s)$ is the qubit's periodic trajectory with period $T_s$, and $f(t')$ is the modulation function ($f(t')=1$ for Ramsey, $f(t')=\pm 1$ for echo). For nonuniform noise, where the noise character depends on the absolute position, it is necessary to consider the full spatiotemporal correlation function $C(x,x',t_1,t_2) = \langle \delta\omega(x,t_1) \delta\omega(x',t_2) \rangle$. To make the problem tractable, we propose a coarse-grained model (See Fig.~\ref{fig:coarse_grained} for illustration), which we describe next.
\subsection{Coarse-Grained Model}

We discretize the space into regions centered at positions $x_i$. Each region has width $\Delta x = d/M$, where $d$ is the total shuttling distance and $M$ is the number of regions. Assuming constant shuttling speed, we also discretize a single period into $2M$ segments of duration $\Delta t$ ($T_s = 2M\Delta t$). The accumulated phase over $\tau = NT_s$ is:
\begin{equation}
    \delta \phi(\tau) \approx \sum_{n=0}^{N-1}\sum_{i=1}^{M} \
    \int_0^{T_s} \text{d}t' \, \delta\tilde\omega(x_i, t' + nT_s) f_i(t' + nT_s),
\end{equation}
where $\delta \tilde \omega(x_i,t')\approx \int \rho(x-x_i) \delta \omega(x,t') \, \text{d}x$ and $f_i(t')= 1$ when the electron is in region $i$ and is $0$ otherwise. The decoherence factor $\chi(\tau) = \frac{1}{2}\langle \delta \phi(\tau)^2 \rangle$ is:
\begin{equation}
    \chi(\tau) = \frac{1}{2} \sum_{i,j} \int_{-\infty}^{\infty} \frac{d\omega}{2\pi} S(x_i, x_j, \omega) \, \mathcal{F}_{ij}(\omega, N),
    \label{eq:chi_filter_general}
\end{equation}
where $S_{ij}(\omega) = S(x_i, x_j, \omega)$ is the cross-spectral density and $\mathcal{F}_{ij}(\omega, N) = \text{Re}\left[ F_i(\omega, N) F_j^*(\omega, N) \right]$ is the total spatiotemporal filter, which depends on the experiment.

\subsection{Ramsey Sequence}
Due to periodicity of the segment function $f_i(t+nT_s) = f(t)$, its Fourier transform $F_i(\omega, N)= \sum_{n=0}^{N-1} \int_{0}^{T_s} \text{d}t' f_i(t'+nT_s) e^{-i\omega (t'+nT_s)}$ decomposes into a single-period filter $F_i^{(1)}(\omega) $ and a periodic comb $G_R(\omega)$:
\begin{equation}
    F_i(\omega, N) = \left( \int_0^{T_s} dt \, f_i(t) e^{-i\omega t} \right) \left( \sum_{n=0}^{N-1} e^{-i\omega nT_s} \right) \equiv F_i^{(1)}(\omega) \cdot G_N(\omega) \;.
\end{equation}
As a result, the total filter $\mathcal{F}_{ij}$ factorizes:
\begin{equation}
    \mathcal{F}_{ij}(\omega, N) = \text{Re}\left[ F_i^{(1)}(\omega) F_j^{(1)*}(\omega) \right] \cdot |G_N(\omega)|^2 \equiv W_{ij}(\omega) \cdot |G_N(\omega)|^2 \;.
\end{equation}
The comb filter $|G_N(\omega)|^2$ is:
\begin{equation}
\label{eq:G_ramsey}
    |G_N(\omega)|^2 = \frac{\sin^2(N\omega T_s/2)}{\sin^2(\omega T_s/2)} \approx N^2 \sum_{k=-\infty}^{\infty} \sinc^2\left(\frac{(\omega - k\omega_s)NT_s}{2}\right)
\end{equation}
where $\omega_s = 2\pi/T_s$. The inter-period function $W_{ij}(\omega) = \int_0^{T_s} \int_0^{T_s} \text{d}t\text{d}t' f_i(t) f_j(t) \cos(\omega(t-t'))$ can be simplified in a coarse-grained model where the electron motion is approximated as a stepwise trajectory. The total time $T_s$ is divided into $2M$ discrete time steps of duration $\Delta t$.

The segment function $f_i(t)$, which tracks when the electron is in region $i$, can be written as a sum of boxcar functions $\Pi(t)$:
\begin{equation}
    f_i(t) \approx \sum_{m=1}^{2M} \delta_{i, k_m} \, \Pi\left(\frac{t - t_m}{\Delta t}\right),
\end{equation}
where $k_m$ is the region index at time step $m$, $t_m = (m-1/2)\Delta t$ is the center of the time step, and $\Pi(x) = 1$ for $|x|<1/2$ and $0$ otherwise. Substituting this into the definition of $W_{ij}(\omega) = \text{Re} [F_i^{(1)}(\omega) F_j^{(1)*}(\omega)]$, the Fourier transform separates into two distinct components: the Fourier transform of the boxcar shape and the phase factors arising from the time shifts $t_m$:
\begin{align}
    W_{ij}(\omega) &= \text{Re} \left[ \left( \sum_{m=1}^{2M} \delta_{i, k_m} e^{-i\omega t_m} \tilde{\Pi}(\omega) \right) \left( \sum_{p=1}^{2M} \delta_{j, k_p} e^{+i\omega t_p} \tilde{\Pi}^*(\omega) \right) \right] \notag \\
    &= |\tilde{\Pi}(\omega)|^2 \sum_{m,p} \delta_{i, k_m} \delta_{j, k_p} \text{Re}\left[ e^{-i\omega(t_m - t_p)} \right] \;.
\end{align}
Using the known Fourier transform of a boxcar, $|\tilde{\Pi}(\omega)|^2 = (\Delta t)^2 \text{sinc}^2(\omega \Delta t / 2)$, and noting that $t_m - t_p = (m-p)\Delta t$, we obtain the factorized form:
\begin{equation}
\label{eq:Wij_ramsey_intuitive}
    W_{ij}(\omega) = \underbrace{\left( \Delta t \operatorname{sinc}\frac{\omega \Delta t}{2} \right)^2}_{\text{Pulse Shape (Low-pass)}} \cdot \underbrace{\sum_{m=1}^{2M} \sum_{p=1}^{2M} \delta_{i, k_m} \delta_{j, k_p} \cos(\omega(m-p)\Delta t)}_{\text{Structure Factor (Interference)}} \;.
\end{equation}
This factorization offers a clear physical intuition: the term $|\mathcal{F}_R|^2 = (\Delta t \sinc \omega \Delta t/2)^2$ arises from the Ramsey-like experiment in each segment, acting as a low-pass filter that suppresses noise frequencies $\omega \gg 1/\Delta t$. The second term ${K_{ij}(\omega, \Delta t) =\sum_{m=1}^{2M} \sum_{p=1}^{2M} \delta_{i, k_m} \delta_{j, k_p} \cos(\omega(m-p)\Delta t) }$ represents the interference pattern created by the specific sequence of visits to spatial regions $i$ and $j$.

\subsection{Example of $K_{ij}(\omega, \Delta t)$ Calculation}
It is instructive to consider a concrete example of calculating the structure factors $K_{ij}(\omega, \Delta t)$ for specific shuttling sequences, shown in Fig.~\ref{fig:coarse_grained}. For illustration let's consider a shuttling sequence that visits four regions labeled A, B, C, and D in an 8-segment period. Here $2M=8$, $T_s=8\Delta t$. The sequence is $k_m = \{A, B, C, D, D, C, B, A\}$. Each region is visited twice ($N_i=2$). We can compute the following structure factors explicitly:
\begin{itemize}
    \item $K_{AA}(\omega, \Delta t)$: Visits at $m \in \{1, 8\}$. Pairs are (1,1), (8,8), (1,8), (8,1).
    $$ K_{AA}(\omega, \Delta t) = \cos(0) + \cos(0) + \cos(-7\omega\Delta t) + \cos(7\omega\Delta t) = 2 + 2\cos(7\omega\Delta t) $$
    \item $K_{DD}(\omega, \Delta t)$: Visits at $m \in \{4, 5\}$. Pairs are (4,4), (5,5), (4,5), (5,4).
    $$ K_{DD}(\omega, \Delta t) = \cos(0) + \cos(0) + \cos(-\omega\Delta t) + \cos(\omega\Delta t) = 2 + 2\cos(\omega\Delta t) $$
    \item  $K_{AB}(\omega, \Delta t)$: Visits at $m \in \{1, 8\}$ for A, $p \in \{2, 7\}$ for B. Pairs: (1,2), (1,7), (8,2), (8,7).
    $$ K_{AB}(\omega, \Delta t) = \cos(-\omega\Delta t) + \cos(-6\omega\Delta t) + \cos(6\omega\Delta t) + \cos(\omega\Delta t) = 2\cos(\omega\Delta t) + 2\cos(6\omega\Delta t) $$
\end{itemize}

\subsection{DC and AC Contributions}
The comb filter $|G_N(\omega)|^2$ has sharp peaks at harmonics of the shuttling frequency $\omega_k = k\omega_s$. While for $|k| > 0$ this can be approximated as delta functions, the $k=0$ peak at $\omega=0$ requires additional care, if the noise spectrum has significant low-frequency weight. We therefore separate the decoherence into DC ($k=0$) and AC ($k \neq 0$) contributions, such that in total:
\begin{equation}
    \chi(\tau) = \chi_{DC}(\tau) + \chi_{AC}(\tau) \;.
\end{equation}

For the low-frequency contribution we generalize the two-point DC approximation from Appendix \ref{app:two_point} to multiple regions. We assume the cross-spectrum can be written as $S_{ij}(\omega) = r_{ij} \sqrt{S_{ii}(\omega) S_{jj}(\omega)}$, where $r_{ij}$ is the normalized correlation coefficient between regions $i$ and $j$. By taking the limit of $\omega \to 0$ in Eq. \eqref{eq:chi_filter_general}, we evaluate the structure factor at $\cos(\omega(m-p)\Delta t) \approx 1$, which gives $K_{ij}(0, T_s) = N_iN_j$. We also take $|G_N(\omega)|_{\omega \to 0}^2 = N^2$, but keep $|\mathcal{F}_R(\omega, \Delta t)|^2 = (\Delta t \sinc \omega \Delta t/2)^2$ general, as $\mathcal{F}_R(\omega, \Delta t)$ is much wider than the $\omega=0$ peak of $G_N(\omega)|_{\omega \to 0}$. With those assumptions and for $1/\omega^\beta$ spectra, we find:
\begin{equation}
\label{eq:chi_shuttledc}
    \chi_{DC}(\tau) \approx \sum_{i,j} r_{ij} \sqrt{ \chi_i(\tau/M) \chi_j(\tau/M) } = \sum_{ij} r_{ij} \left(\frac{\tau/M}{T_{2,i}^*}\right)^{\alpha_i/2} \left(\frac{\tau/M}{T_{2,j}^*}\right)^{\alpha_j/2}
\end{equation}
where $\chi_i(\tau/M) = (\tau/MT_{2,i}^*)^{\alpha_i}$ is the stationary decoherence function for region $i$ evaluated at the effective time $\tau/M$. We note that the effective time $\tau/M$ arises from the assumption that the qubit visits each of the $M$ regions exactly twice per period (forward and backward pass), resulting in a total time fraction of $2/2M = 1/M$.

On the other hand, the AC contribution arises from harmonics $k \neq 0$. As the spectral density is typically flat, the comb filter has delta functions at $\omega_k = k\omega_s$, allowing us to write:        
\begin{equation}
        \chi_{AC}(\tau) \approx \left(\frac{N}{2T_s}\right) \sum_{k \neq 0} \sum_{i,j} S_{ij}(\omega_k) W_{ij}(\omega_k) \;.
\end{equation}

Finally, while the choice of $M$ segments is arbitrary, it should reflect a meaningful spatial scale over which the noise properties change. In practice, $\Delta x$ should be larger than the electron wavefunction extent, but smaller than the characteristic correlation length of the noise. For large separations between the segments the model is expected to interpolate between the missing probing points, which should lead to reasonable estimates of the overall decoherence, unless the noise has large spatial variations at scales smaller than $\Delta x$.
\subsection{Numerical verification}
\begin{figure}[htb!]
    \includegraphics[width=\textwidth]{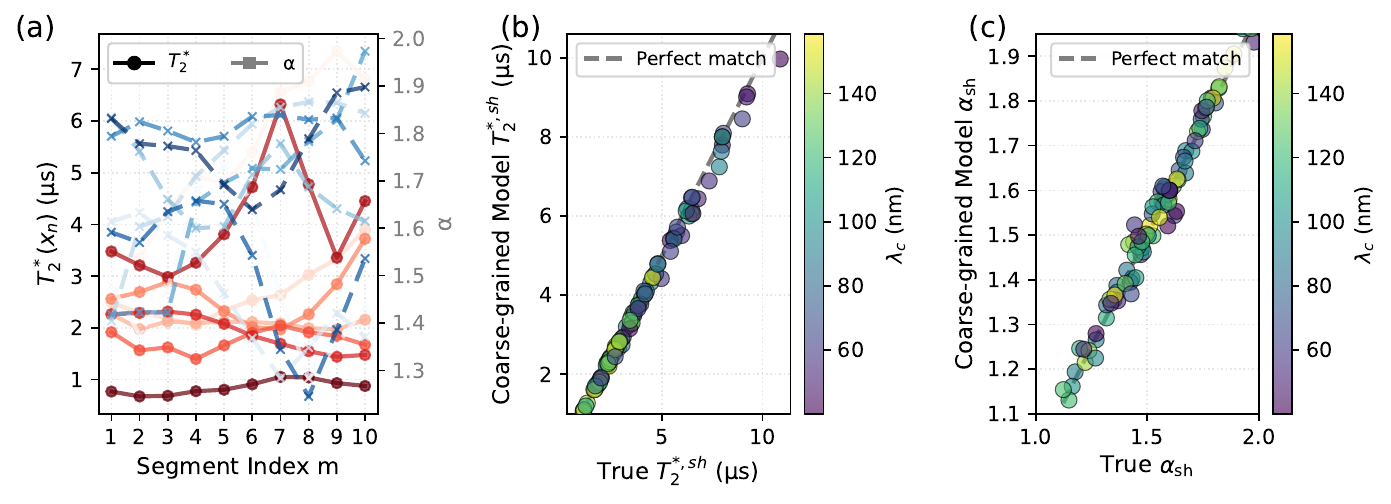}
    \caption{Validation of the coarse-grained model for periodic shuttling through a spatially varying noise landscape. Typical noise realizations  are depicted in panel (a). Panel (b) compares the extracted and true $T_{2}^{*,sh}$ and panel (c) compares the extracted and true $\alpha_{sh}$ from fits to the decoherence computed using the full continuous noise landscape and the coarse-grained model. The color in panels (b) and (c) represents the correlation length $\lambda_c$.}
\label{fig:bucket_model_validation}
\end{figure}

To validate the coarse-grained model, we perform numerical simulations of a qubit shuttled periodically through a spatially varying noise landscape. We generate a 1D noise profile $\delta\omega(x,t)$ by simulating 40 Ornstein-Uhlenbeck processes with a distribution of correlation times generating $1/\omega^\beta$ noise \cite{krzywda_adiabatic_2020}. At each point of the conveyor we sum the contributions from all fluctuators to obtain the total noise $\delta\omega(x,t) = \sum_i v_i(x) \xi_i(t)$ with $v_i(x) = \exp(-|x - x_i|/\lambda_{c})$. Sampled $T_2^*(x)$ and $\alpha(x)$ in each segment for several realization are shown in Fig.~\ref{fig:bucket_model_validation} (a).

The main goal is to validate the approach from the main text, which was used on the experimental data, where the ground truth is not known. Here by simulating the system, we can compare both coarse-grained and DC-approximation models against the ground truth provided by the fully simulated result. For each realization we sample the correlation times, correlation length $\lambda_{c}$ and the locations of the fluctuators over a domain of width $2d$ (twice the shuttling distance $d$) to ensure the noise bath fully encompasses the qubit trajectory. For each configuration we repeat the simulation over 1000 noise repetitions and compute the decoherence in three ways:
\begin{itemize}
\item Full model: By directly integrating the accumulated phase $\delta \phi(\tau)$ over the continuous noise landscape $\delta\omega(x,t)$.
\item Coarse-grained: By discretizing the noise landscape into $M$ segments according to Eq.~\eqref{eq:chi_filter_general}.
\item DC approximation: By using the simulated stationary $T_2^*(x_n)$ and $\alpha$ parameters at each segments and evaluating the DC approximation for $\chi_{DC}(\tau)$. 
\end{itemize}

The comparison between ground truth and the coarse-grained model is plotted in Fig.~\ref{fig:bucket_model_validation}. By a figure of merit we take the extracted parameters of decoherence function $\chi(\tau)$, i.e. effective $T_2^{*,sh}$ and $\alpha_{sh}$ extracted from the fit. Clearly the coarse-grained model closely matches the true decoherence in regards to both $T_2^{*,sh}$ and $\alpha_{sh}$.

\begin{figure}
    \includegraphics[width=\textwidth]{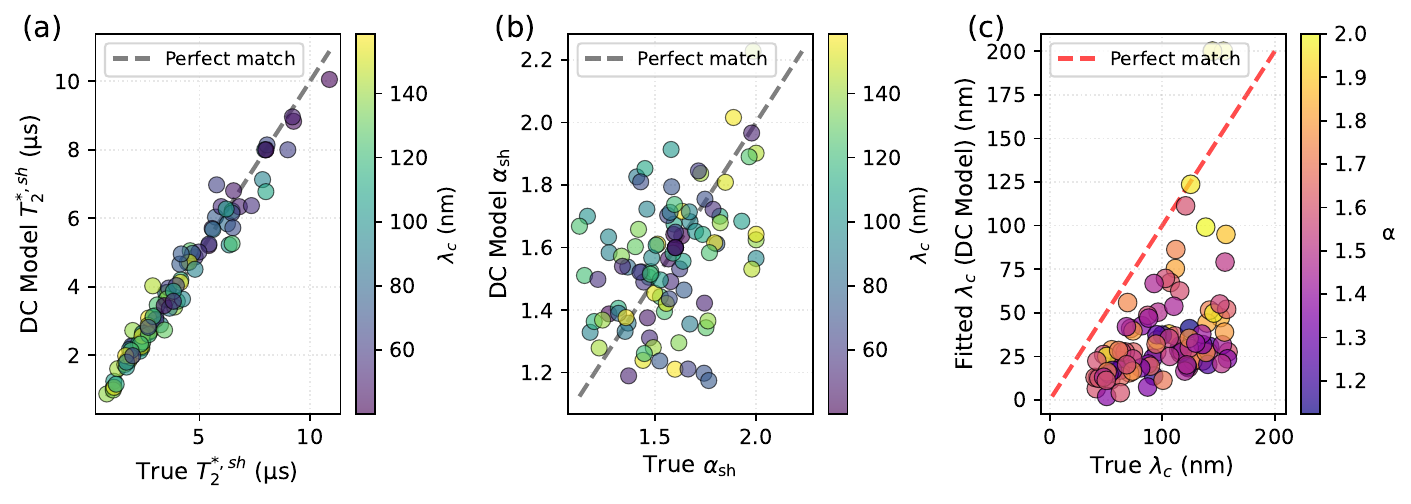}
    \caption{Validation of the DC model for periodic shuttling through a spatially varying noise landscape. We used realizations of the spatiotemporal disorder from Fig.~\ref{fig:bucket_model_validation}. In panel (a) we compare the fitted and true correlation length $\lambda_c$ showing the tendency of DC model to underestimate the true value. This is caused by partial decorrelation of high-frequency noise as indicated by the color gradient - the agreement is stronger for low-frequency noise for which $\alpha \sim 2$. In panel (b) we compare the extracted $T_{2}^{*,sh}$ and in panel (c) $\alpha_{sh}$ with the color indicating the correlation length $\lambda_c$. }
    \label{fig:DC_validation} 
\end{figure}

Next we show the limitation of the DC approximation by comparing extracted $T_2^{*,sh}$ and $\alpha_{sh}$ as well as the correlation coefficient parametrized by the correlation length $\lambda_c$ against the target values from simulation. The procedure amounts to first extracting stationary $T_{2,i}^*$, $
\alpha_i$ in each segment, followed by a fit of $\lambda_c$ in $r_{ij} = \exp(|x_i-x_j|/\lambda_c)$ that minimize the difference between simulated decoherence during shuttling $\chi(\tau)$ and $\chi_{DC}(\tau)$ given by Eq.~\eqref{eq:chi_shuttledc}. The result of this procedure are shown in Fig.~(\ref{fig:DC_validation}), where the agreement in the $T_{2}^{*,sh}$ in panel a, proves the successful fitting procedure. Next, from panel b, we conclude that the decay exponent $
\alpha_{sh}$ from DC approximation is moderately correlated with the true $\alpha_{sh}$. Finally we show that obtained in the procedure correlation length $\lambda_c$ can be treated as a lower bound of true $\lambda_{c}$, which is equivalent of saying that the true value is at least as large as the fitted one. This is a direct consequence of finite temporal correlations. As the electron travels between the regions, the noise can appear not as correlated as it actually is, leading to an underestimate of true $\lambda_{c}$. This is confirmed by observing that the agreement improves for higher $\alpha_{sh}$ (colors in panel c), where this decay of temporal correlations is less pronounced in case of slower noise dynamics. 

Overall, the coarse-grained model provides a reliable framework for modeling decoherence during periodic shuttling through spatially varying noise, while the DC approximation offers a computationally efficient method for estimating decoherence and remaining spatial correlations based on the stationary $T_2^*(x_n)$ and $\alpha(x_n)$ measurements.

\newpage

\,

\newpage

\section{Filtering properties of shuttling and echo sequences}
\label{app:echo_filter}
After introducing the DC and AC contributions in Appendix~\ref{app:shuttling_filter}, we now analyze the filtering properties of periodic shuttling for both Ramsey and echo sequences. Following the previous result, the total decoherence factor is given by:
\begin{equation}
\label{eq:filtering_full}
    \chi(\tau) = \frac{1}{2} \int_{-\infty}^{\infty} \frac{d\omega}{2\pi} \sum_{i,j}  S_{ij}(\omega) \, K_{ij}(\omega, \Delta t)  \, |G_N(\omega)|^2 |\mathcal F_R(\omega,\tau)|^2 \approx \chi_{AC} + \chi_{DC}
\end{equation}
where $S_{ij}(\omega)$ is the cross-spectral density between buckets $i$ and $j$, $K_{ij}(\omega, \Delta t)$ is the interference term between the segments, $|\mathcal F_R(\omega,\tau)|^2$ is the single-segment filter function, both defined in Eq.~\eqref{eq:Wij_ramsey_intuitive}, and $|G_N(\omega)|^2$ is the comb filter due to periodic shuttling (Eq.~\eqref{eq:G_ramsey}) that allows us to split the contributions into a low-frequency part $\chi_{DC}$ and a high-frequency part $\chi_{AC}$. We illustrate the filtering property of the shuttling sequence in Fig.~\ref{fig:shuttling_filter}, where we compare the Ramsey and Hahn-echo sequences.

\subsection{Hahn-echo Sequence}
While for a Ramsey experiment, $G_N(\omega) = \sin^2(N\omega T_s/2)/\sin^2(\omega T_s/2)$ is sensitive to low-frequency noise, an echo sequence can be used to suppress the effect of quasi-static noise. We consider a single echo pulse applied at the midpoint of the total evolution time $t=\tau/2 = (N/2)T_s$. As a result the modulation function for the $i$-th bucket changes sign $f_i(t)=+1$ for $t<\tau/2$ and $f_i(t)=-1$ for $t>\tau/2$ and $f_i(t)= 0$ for times when the electron is not in the bucket $i$. This modifies the inter-period comb filter $G_N(\omega)$ to $G_H(\omega)$:
\begin{equation}
    G_H(\omega) = \sum_{n=0}^{N/2-1} e^{-i\omega nT_s} - \sum_{n=N/2}^{N-1} e^{-i\omega nT_s} = \left( 1 - e^{-i\omega \tau/2} \right) \sum_{n=0}^{N/2-1} e^{-i\omega nT_s}\;.
\end{equation}
\begin{figure}[htb!]
    \includegraphics[width=0.99\textwidth]{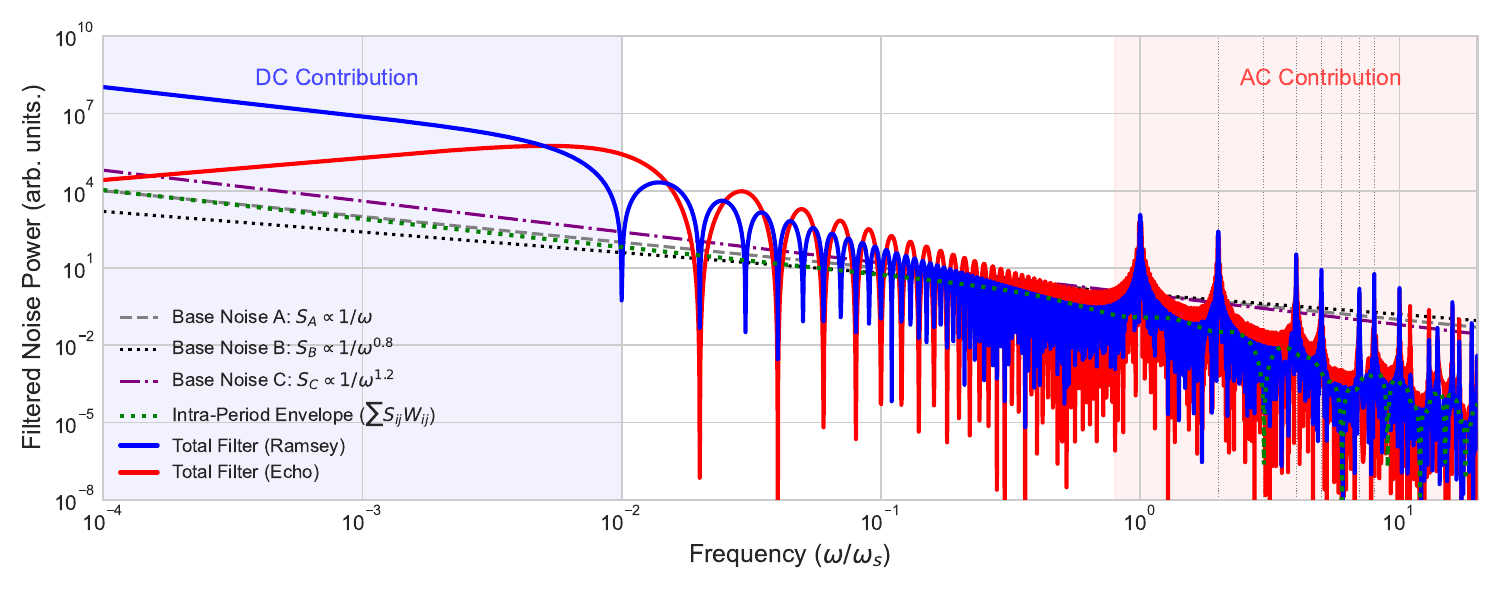}
    \caption{Filtering properties of periodic shuttling sequences. We compare the filtered spectrum in the case of Ramsey (blue) and Hahn-echo (red) sequences during periodic shuttling through a spatially varying noise landscape. For modeling purposes we use three buckets with correlation coefficients $r_{AB} = r_{BC} = 0.4$ and $r_{AC} = 0.2$ with the spectral densities $S_i(\omega) = A_i/\omega^{\beta_i}$ for $i \in \{A, B, C\}$ with $\beta_A = 0.8$, $\beta_B = 1$, and $\beta_C = 1.2$. By the shaded regions, we indicate the parts of the spectrum relevant for the DC the contribution (low-frequency) and the AC contribution (high-frequency). By the green dashed line we show the effect of destructive interference in the interference factor $K_{ij}(\omega, \Delta t)$, which can suppress sensitivity to specific noise frequencies.}
    \label{fig:shuttling_filter}
\end{figure}

The total filter for the echo sequence is $\mathcal{F}_{ij}(\omega, \tau) = W_{ij}(\omega) \cdot |G_H(\omega)|^2$, where the new comb filter is:
\begin{equation}
    |G_H(\omega)|^2 = \left| \frac{(1 - e^{-i\omega \tau/2})^2}{1 - e^{-i\omega T_s}} \right|^2 = 4 \frac{\sin^4(\omega NT_s/4)}{\sin^2(\omega T_s/2)}= 4 \sin^2(\omega NT_s/4)|G_{N/2}(\omega)|^2
\end{equation}
The intra-period weight $W_{ij}(\omega) = |\mathcal F_R(\omega, \Delta t)|^2 K_{ij}(\omega, \Delta t)$ is unchanged (assuming the echo pulse is applied during a stationary interval), as the pulse is applied between periods, not within them. The most pronounced effect of the echo is the suppression of the DC peak at $\omega=0$, as $|G_H(\omega)|_{\omega \to 0}^2 \approx (N/2)^2(\omega \tau)^2/4$, compared to $|G_N(\omega)|_{\omega \to 0}^2 = N^2$ for a Ramsey experiment.

We illustrate this effect in Fig.~\ref{fig:shuttling_filter}, where we plot the effective spectrum (integrand of Eq.~\eqref{eq:filtering_full}) for both Ramsey and Hahn-echo sequences. The echo sequence strongly suppresses the effect of low-frequency noise, while retaining sensitivity to higher-frequency noise at harmonics of the shuttling frequency $\omega_k = k\omega_s$. Both filters have additional modulation from the interference factor $K_{ij}(\omega, \Delta t)$ and the sinc$^2$ envelope from the single-segment filter $|\mathcal{F}_R(\omega, \Delta t)|^2$.

\newpage

\section{Effective Model of Dressed-State Shuttling \label{app:driven_qubit}}

\begin{figure}[h!]
    \centering
    \includegraphics[width=0.99\linewidth]{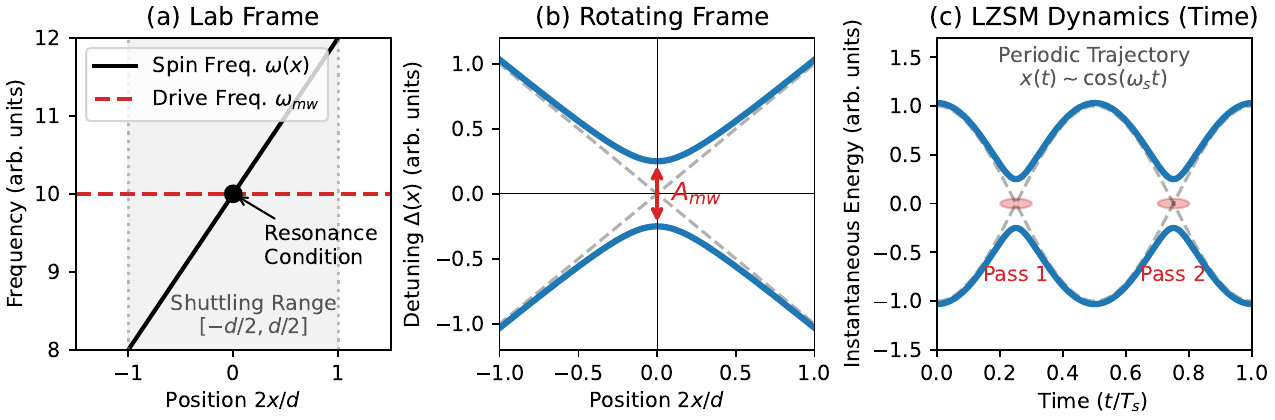}
    \caption{Effective model of dressed-state shuttling. (a) Energy diagram in the laboratory frame showing the position-dependent qubit frequency $\omega_q(x)$ crossing the fixed microwave drive frequency $\omega_{mw}$. (b) In the rotating frame, the drive opens an avoided crossing (gap $A_{mw}$) at the resonance point. (c) Time evolution of the instantaneous energy levels during periodic shuttling, showing two passages through the avoided crossing per period}
    \label{fig:LZSM_cartoon}
\end{figure}

In this section, we derive the effective Hamiltonian for a qubit undergoing periodic shuttling while driven by a resonant microwave field. We focus on the experimentally relevant regime where the amplitude of the microwave drive is much smaller than the amplitude of spin-splitting modulation. As illustrated in Fig.~\ref{fig:LZSM_cartoon}, we show that in this regime, the system behaves as a dressed qubit undergoing Landau-Zener-Stuckelberg-Majorana interference \cite{shevchenko2010landau} with a renormalized Rabi frequency. While LZSM interferometry has been realized in spin qubit devices before, it relied on coherent tunneling in a double quantum dot, including the $S-T_-$ avoided crossing \cite{Gaudreau_2011} or charge-valley coupling \cite{mi2018landau}. Here for the first time, similar physics is demonstrated with a continuously shuttled mobile spin qubit. 

\subsection{Derivation of the Effective Hamiltonian (LZSM Model)}

We start with the time-dependent Hamiltonian in the lab frame, which includes a microwave drive, a periodic modulation induced by shuttling back-and-forth at angular frequency $\omega_s = 2\pi/T_s$, and a noise term:
\begin{equation}
H(t) = \frac{\Omega_z(t) + \xi(t)}{2}\sigma_z + \frac{\Omega_x(t)}{2}\sigma_x
\end{equation}
where $\Omega_z(t) = \overline \omega_q - A_z \cos(\omega_s t)$, with the average qubit frequency $\overline \omega_q = (1/T_s) \int_0^{T_s} \text{d}t \int \text{d}x\rho(x-x(t))\omega_q(x,t)$ and the amplitude of modulation of the spin-splitting $2A_z$, which can be related to the gradient of the longitudinal magnetic field through $A_z = \nabla B_\parallel d/2$, for a constant gradient $\nabla B_\parallel$, where $d$ is the shuttling distance. $\Omega_x(t) = 2A_{mw}\cos(\omega_{mw} t)$ represents the external microwave drive with amplitude $2A_{mw}$ and frequency $\omega_{mw}$. The $\xi(t)$ is the wavefunction-averaged noise term $\xi(t) = \int \text{d}x \rho(x-x(t)) \delta\omega(x, t)$.

\subsubsection{Step 1: Rotating Wave Approximation (RWA)}
We first move to a frame rotating with the microwave drive at $\omega_{mw}$ using the transformation $U_1 = e^{i (\omega_{mw} t / 2) \sigma_z}$. Applying the RWA \cite{AllenEberly1987} (neglecting terms oscillating at $2\omega_{mw} \gg A_{mw}$) simplifies the Hamiltonian to
\begin{equation}
\label{eq:H_rwa}
H_{RWA}(t) = \frac{(\overline \omega_q - \omega_{mw}) - A_z \cos(\omega_s t) + \xi(t)}{2}\sigma_z + \frac{A_{mw}}{2}\sigma_x \equiv \frac{\Delta (t)}{2} \sigma_z + \frac{A_{mw}}{2}\sigma_x \;.
\end{equation}
This is the Hamiltonian for a ``dressed qubit", where $A_{mw}$ acts as a static transverse field (the new time-averaged quantization axis when $\overline \omega_q = \omega_{mw}$) and the shuttling term $A_z$ acts as a drive amplitude in this basis that translates to time-depedent detuning $\Delta (t)$. If additionally $A_z \gg A_{mw}$ the system undergoes Landau-Zener-Stückelberg-Majorana (LZSM) interferometry.

In the wider context of qubit shuttling, This Hamiltonian could equivalently describe repeated shuttling of electron charge between two tunnel-coupled quantum dots, with $A_{mw}$ representing the tunnel coupling and $A_z$ the detuning modulation amplitude.

\subsubsection{Step 2: Floquet Theory \& Multi-Photon Resonance}
In the absence of noise, $H_{RWA}(t)$ is periodic in time with period $T_s = 2\pi / \omega_s$. According to Floquet theory, multi-photon resonances occur when the static detuning matches a multiple of the shuttling (drive) frequency:
\[
\Delta = \overline{\omega}_q - \omega_{MW} \approx k\omega_s
\]
where $k$ is an integer.

We now follow a standard derivation of LZSM Hamiltonian in the Floquet space \cite{shevchenko2010landau} and move to a second rotating frame $U_2 = e^{i \frac{A_z}{\omega_s} \sin(\omega_s t) \sigma_z / 2}e^{i k\omega_s \sigma_z / 2}$, transforming to the dressed-state (Floquet) basis:
\begin{equation}
H_{F}(t) = U_2 H_{RWA}(t) U_2^\dagger - i U_2 \dot{U}_2^\dagger = \frac{(\overline \omega_q- \omega_{MW} - k\omega_s) + \xi(t)}{2}\sigma_z + \frac{A_{mw}}{2} \left[\sigma_+ e^{i \frac{A_z}{\omega_s} \sin(\omega_s t)\sigma_z + ik\omega_s\sigma_z} + \sigma_- e^{-i \frac{A_z}{\omega_s} \sin(\omega_s t)\sigma_z - ik\omega_s\sigma_z}\right].
\end{equation}

We apply the second rotating wave approximation (RWA) around the $k$-th resonance, motivated by the fact that $A_{mw} \ll \omega_s$, i.e. the shuttling frequency is much faster than the Rabi frequency resulting from the microwave drive. As a result, the transverse terms not in resonance with $k\omega_s$ are averaged out, and in the Floquet basis the effective Hamiltonian becomes time-independent:
\begin{equation}
\label{eq:hk}
H_{k} = \frac{1}{T} \int_0^T H_F(t) \text{d}t = \frac{(\overline \omega_q- \omega_{MW} - k\omega_s) + \xi_T}{2}\sigma_z + \frac{A_{mw}}{2} \ J_k\left(\frac{A_z}{\omega_s}\right) \sigma_x
\end{equation}

where we used the Jacobi-Anger expansion ($e^{i \beta \sin(\omega_qt)} = \sum_n J_n(\beta) e^{in\omega_qt}$) with Bessel Functions of the first kind $J_n(\beta)$. Note that strictly speaking, for odd $k$, the transverse coupling operator rotates to $\sigma_y$ due to the Bessel function property $J_{-k}(\beta) = (-1)^k J_k(\beta)$, but for the $k=0$ resonance focused on here, the coupling remains proportional to $\sigma_x$. Clearly the time integral selects the $k$-th harmonic, as $(1/T_s)\int_0^{T_s} e^{i(n-k)\omega_s t} dt = \delta_{n,k}$. Finally, the effective noise term also becomes time-independent:
\[
\xi_T = \frac{1}{T_s} \int_0^{T_s} dt \int dx \rho(x-x(t)) \delta\omega_q(x, t),
\]
as it is averaged over a full shuttling period.

\begin{figure}[htb!]
    \centering
    \includegraphics[width=0.99\linewidth]{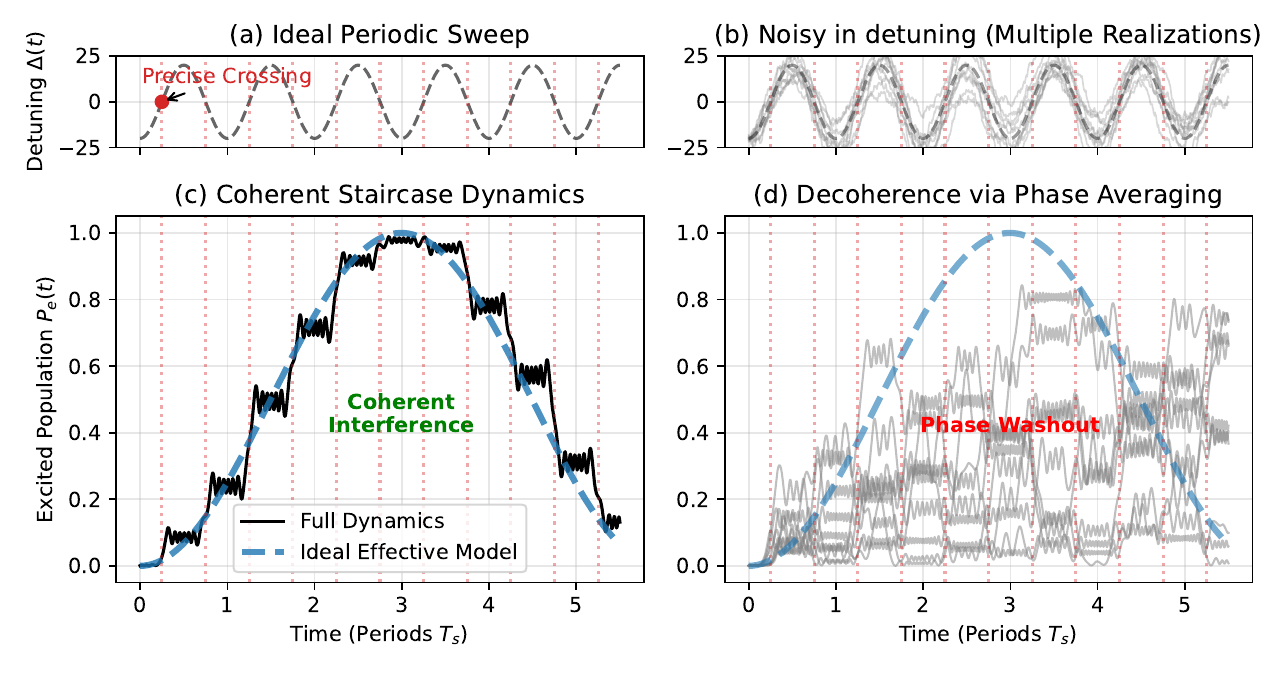}
    \caption{Simulation of LZSM dynamics and decoherence. (a) Ideal periodic detuning sweep $\Delta(t)$ across the resonance. (b) Detuning trajectories including multiple realizations of quasistatic noise. (c) Coherent "staircase" population evolution in the noiseless limit, demonstrating agreement between the full dynamics (black) and the effective model (dashed blue). (d) Population evolution in the presence of noise, showing how phase averaging leads to signal washout and decoherence.}
    \label{fig:LZSM_interference}
\end{figure}

\subsection{Landau-Zener Stuckelberg Majorana picture}

Above, we derived a framework where the shuttled spin is effectively driven through avoided crossings induced by the external microwave drive. Transitions between spin eigenstates arise from Landau-Zener-Stückelberg-Majorana (LZSM) interference. As illustrated in Fig.~\ref{fig:LZSM_interference}~(a), the detuning is swept periodically across the resonance.

When on resonance, the interference between consecutive crossings becomes constructive, resulting in a coherent accumulation of probability amplitude (Fig.~\ref{fig:LZSM_interference}~(c)). While the full Hamiltonian captures the fast oscillations between consecutive transitions (black solid line), the effective Hamiltonian accurately captures the envelope of the dynamics (blue dashed line).

In the presence of noise, this constructive interference is disrupted. Fluctuations in the detuning (Fig.~\ref{fig:LZSM_interference}~(b)) introduce random phase shifts. Consequently, phase averaging leads to signal washout and decoherence, as shown in Fig.~\ref{fig:LZSM_interference}~(d).

\subsection{Numerical Verification of Effective Hamiltonian}

\begin{figure}[htb!]
    \centering
    \includegraphics[width=0.99\linewidth]{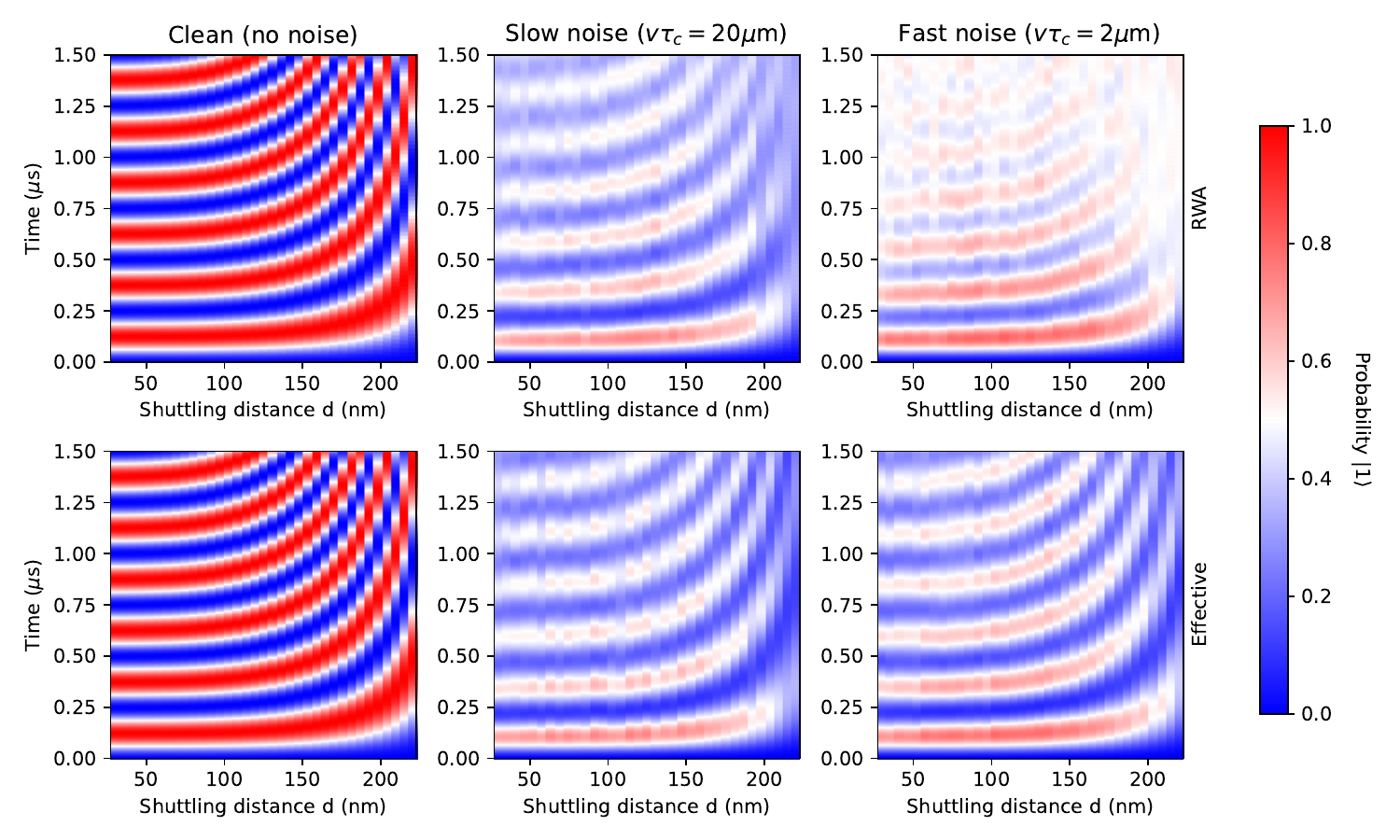   }
    \caption{Comparison of the time evolution of the qubit state under the full RWA Hamiltonian and the Effective Hamiltonian.
The color scale indicates the qubit population expectation value $\langle \sigma_z \rangle$, ranging from ground state (blue) to excited state (red). 
(Top Row) Numerical simulation of the time evolution using the full $H_{\text{RWA}}$ (Eq.~\eqref{eq:H_rwa}) for (left) a noiseless system, (center) slow noise with correlation time $\tau_c = 100\,$ns, and (right) fast noise with $\tau_c = 10\,$ns. 
(Bottom Row) Corresponding simulations using the derived Effective Hamiltonian $H_{k,\text{eff}}$. 
The shuttling velocity is fixed at $v=10\,$m/s. The strong visual agreement between the top and bottom rows verifies that the effective Hamiltonian with period-averaged noise $\xi_T$ quantitatively captures the decoherence dynamics of the full system.}
    \label{fig:Hrwa}
\end{figure}

We now numerically verify the validity of the effective Hamiltonian by comparing its predictions to simulations of the $H_{RWA}$ given by Eq.~\eqref{eq:H_rwa}. We simulate full time evolution of the qubit state under $H_{RWA}$(t) in presence of spatiotemporal Ornstein-Uhlenbeck process similar to the one used in \cite{zhangPRB_2025}. We fix the correlation length to $\lambda_c = 100$~nm and investigate two different correlation times, corresponding to slow noise setup $\tau_c = 100$~ns and fast noise regime $\tau_c = 10$~ns. To stay close to experimental conditions we use fixed velocity $v = 10$~m/s, such that the $\omega_s$ is distance-dependent. 

The results are shown in Fig.~\ref{fig:Hrwa}, where an agreement between effective and RWA Hamiltonian is visible in the noiseless and quasistatic case. The discrepancy in the fast noise regime is caused the dissipative evolution that causes the transition between the ground and excited state. While this shows limitation of the effective Hamiltonian and quasistatic noise approximation, in principle it can be modeled by solving the master equation in the effective space with properly defined excitation and relaxation operators. However, this treatment is outside of the scope of this paper. In general, we conclude that effective Hamiltonian approximation with period-averaged noise is sufficient to qualitatively explain decoherence of a periodically shuttled driven qubit undergoing LZSM interference.

\newpage
\section{Coherence Limits in the Driven Regime}
\label{app:coherence_in_driven}
We finally perform a more qualitative analysis of decoherence of a driven system and illustrate the trade-off associated with an increasing shuttling distance, which on one hand allows for more noise averaging but on the other hand can reduce the shuttling frequency.
\subsection{Noise in the Floquet Basis}
We take the effective Hamiltonian in the doubly-rotating frame, and use the fact that the noise is effectively quasistatic, i.e. what contributes to decoherence is the time-independent part $\xi_T$. Concentrating on the resonance, the presence of the noise has a two-fold effect: on the one hand it tilts the rotation axis $n(t) = (\Omega_{R}, 0, \xi_T)$, and on the other hand it leads to fluctuations in the oscillation frequency $\Omega(t) = \sqrt{\Omega_{R}^2+ \xi_T^2}$. As typically the noise is small, i.e. $\xi_T \ll \Omega_{R}$, the correction to the rotation axis can be neglected, and the dominant error is due to the fluctuations of the oscillation frequency:
\begin{equation}
    \Omega(t) = \sqrt{\Omega_{R}^2 + \xi_T^2} \approx \Omega_{R} + \frac{\xi_T^2}{2\Omega_{R}}.
\end{equation}
    which leads to dephasing of the oscillations, i.e.
\begin{equation}
    W(t_f) = |\langle \sigma_z(t) - i\sigma_y(t)\rangle|^2 = \bigg|\bigg\langle \exp\left(-i\int_0^{t_f} \frac{\xi_T^2}{2\Omega_{R}}\text{d}t\right)\bigg\rangle\bigg|.
\end{equation}
Such a contribution can be computed analytically for Gaussian quasistatic noise, i.e. when the noise is constant over the time of interest, with probability distribution $p(\xi) = \frac{1}{\sqrt{2\pi}\sigma} \exp\left(-\frac{\xi^2}{2\sigma^2}\right)$, leading to:
\begin{equation}
\label{eq:W_rabi_floquet}
    W(t_f) = \bigg|\frac{1}{\sqrt{2\pi}\sigma} \int \text{d}\xi_T \exp\left( - \frac{\xi_T^2}{2\sigma^2} - \frac{i \xi_T^2 t_f}{2 \Omega_{R}}\right)\bigg| = \left(1 + \frac{\langle \xi_T^2\rangle^2 t_f^2}{\Omega_{R}^2}\right)^{-1/4},
\end{equation}
and resulting in a polynomial decay of the coherence with the time $t_f$, which is consistent with the effectively transverse nature of the noise during Rabi oscillations \cite{ramon2022qubit, dobrovitski2009decay}. However, in any realistic experiment the overall loss of coherence is a compound effect of many sources, including weaker but exponential or gaussian contributions, which typically prevent measuring long tails of polynomial decay. That is why, for the fitting procedure, it is more convenient to look at the leading order correction to the coherence, given by:
\begin{equation}
    W(t_f) \approx 1 - \frac{\langle\xi_T^2\rangle^2 t_f^2}{4\Omega_R^2},
\end{equation}
and define the Rabi decay time $T_R^{sh}$ through $\langle \xi_T^2 \rangle T_R^{sh}/2\Omega_R = 1$. It allows us to compare theoretical decay times with numerically extracted fits, that in the limit of large times are dominated by other decoherence mechanisms with an exponential or gaussian envelope.
 As a result we can write coherence time as:
\begin{equation}
\label{eq:app_Trabi}
    T_R^{sh} = \frac{2\Omega_R}{\langle \xi_T^2\rangle},
\end{equation}
which shows that the decay of Rabi oscillations depends on the frequency $\Omega_R$ and effective noise amplitude $\langle \xi_T^2\rangle$. Below, we relate both of them to the experimental parameters, showing non-trivial dependence on the shuttling distance $d$.

\subsection{Effective Rabi Frequency}
\begin{figure}[htb!]
    \centering
    \includegraphics[width=1\linewidth]{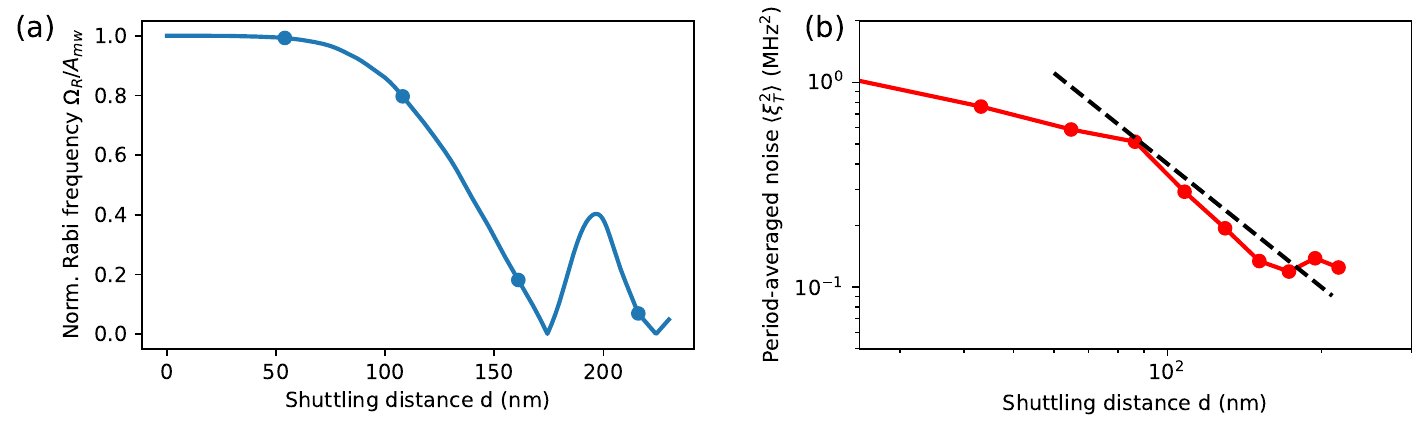}
    \caption{Reconstructed parameters of Effective Hamiltonian (a) The Frequency profile related to the Bessel function. (b) Upper bound of the period-averaged noise $2/[T_s T_{2}^{CPMG,sh}]$ using data from the CPMG shuttling experiment. The dashed line indicates $\propto d^{-2}$.}
    \label{fig:Om_xit}
\end{figure}

As shown in Appendix~\ref{app:driven_qubit}, the parameters of the effective Hamiltonian can be expressed in physically relevant parameters like the magnetic field gradient, shuttling distance and speed. Following the derivation above, the effective Rabi frequency reads $\Omega_R = A_{mw} J_k(\beta)$, where the argument of the Bessel function of the first kind is given by:
\[
\beta = \frac{A_z}{\omega_s} = \frac{d}{2\pi v}\int_{x_0}^{x_d}\nabla\omega_q(x) - \overline{\omega}_q \,dx \approx \frac{\nabla \omega_qd^2}{2\pi v}.
\]
The last approximation was taken under the assumption of a constant gradient $\nabla \omega_q$. We concentrate on the 0-th order resonance $k=0$, for which $\Omega_R \propto J_0(\beta)$. From the properties of Bessel functions, for $\beta \ll 1$, $J_0(\beta) \approx 1 - \beta^2/4$, and in the limit of small $\beta$ we effectively recover the non-shuttling Rabi frequency $\Omega_R \approx A_{mw}$. However, as $\beta$ increases (either by increasing the gradient or shuttling distance, or decreasing the shuttling speed), the Rabi frequency is suppressed, illustrated in Fig.~\ref{fig5_dressed} of the main text. To explain the origin of the measured $\Omega_R$, we convert the extracted the shuttling speeds, and the frequency profile $\omega(x)$ into the argument $\beta$, and in Fig.~\ref{fig:Om_xit}a plot the Rabi frequency $\Omega_R$ normalized to the Rabi drive amplitude $A_{mw}$. The points mark the distances used in the main text, qualitatively confirming two pairs of similar $\Omega_R$ at different $d$.

\subsection{Effective Noise Amplitude $\langle\xi_T^2\rangle$}
Finally, we attempt to relate the period-averaged noise amplitude to a decoherence function measured in the undriven shuttling experiment. By definition, the period-averaged noise variance relates to the decoherence exponent $\chi_{sh}(T_s) \equiv \frac{1}{2}\langle \delta\phi^2(T_s)\rangle$ as:
\begin{equation}
    \langle \xi_T^2 \rangle = \bigg\langle \bigg(\frac{1}{T_s}
    \int_0^{T_s}
    dt \int dx \rho(x-x(t)) \delta\omega(x, t)  \bigg)^2\bigg\rangle \approx \frac{2\chi_{sh}(T_s)}{T_s^2}.
\end{equation}
For the considered distances and shuttling speeds, the single-period duration is much shorter than the measured decoherence time ($T_s = 2d/v < 40\,\text{ns} \ll T_2^{*,sh}$). On such short timescales, the contribution from high-frequency noise can dominate over low-frequency noise. This is evident from the short-time expansion of the coherence function $W(T_s) = \exp[-\chi(T_s)]$:
\begin{equation}
    W(T_s) \approx 1 - \frac{T_s}{T_{\text{fast}}} - \frac{1}{2}\left(\frac{T_s}{T_{\text{slow}}}\right)^2 \approx 1 - \frac{T_s}{T_{\text{fast}}},
\end{equation}
where the linear term (associated with high-frequency exponential decay) can outweigh the quadratic term (low-frequency Gaussian decay) even if $T_{\text{fast}} > T_{\text{slow}}$. 

Next, we establish a bound for the high-frequency timescale $T_{\text{fast}}$. Since dynamical decoupling (DD) sequences filter out low-frequency noise, the coherence time measured in a DD-protected shuttling experiment, $T_{2}^{CPMG,sh}$, approaches the high-frequency limit. Using the fact that $T_{2}^{CPMG,sh}(d) \leq T_{\text{fast}}$ (as experimental imperfections and residual noise limit the measured time below the theoretical white-noise limit), we can use the measured value to place a conservative upper bound on the noise amplitude:
\begin{equation}
    \langle \xi_{T}^2(d) \rangle \approx \frac{2}{T_s}\left( \frac{1}{T_{\text{fast}}} \right) \leq \frac{2}{T_s T_{2}^{CPMG,sh}(d)} = \frac{v}{d\,T_{2}^{CPMG,sh}(d)}.
\end{equation}
This result, plotted in Fig.~\ref{fig:Om_xit} provides a functional form for the effective noise power experienced by the qubit during a single shuttling period. Notably, while the variance $\langle \xi_T^2 \rangle$ scales linearly with velocity $v$ (due to the reduced averaging time of the stochastic noise field), the total accumulated phase error $\langle \delta\phi^2 \rangle \propto \langle \xi_T^2 \rangle T_s^2$ scales as $1/v$, confirming that faster shuttling remains advantageous for fidelity. Similarly, for the constant velocity case, the accumulated phase error increases with distance, $\langle \delta\phi^2 \rangle \propto d/T_{2}^{CPMG,sh}(d)$, as long as the high-frequency coherence time scales sublinearly with distance.

\newpage

\,
   \section{Variability of high-field data \label{app:consistency}}

In this section, we discuss the consistency of the data against parameter changes, in particular the relation between the voltage applied to the conveyor belt and the observable quantities: coherence time and noise correlations. For simplicity, we concentrate on stationary and two-point data, but in principle, a similar analysis can be performed for the shuttling data as well.

\begin{figure}[htb!]
    \centering
    \includegraphics[width=0.88\linewidth]{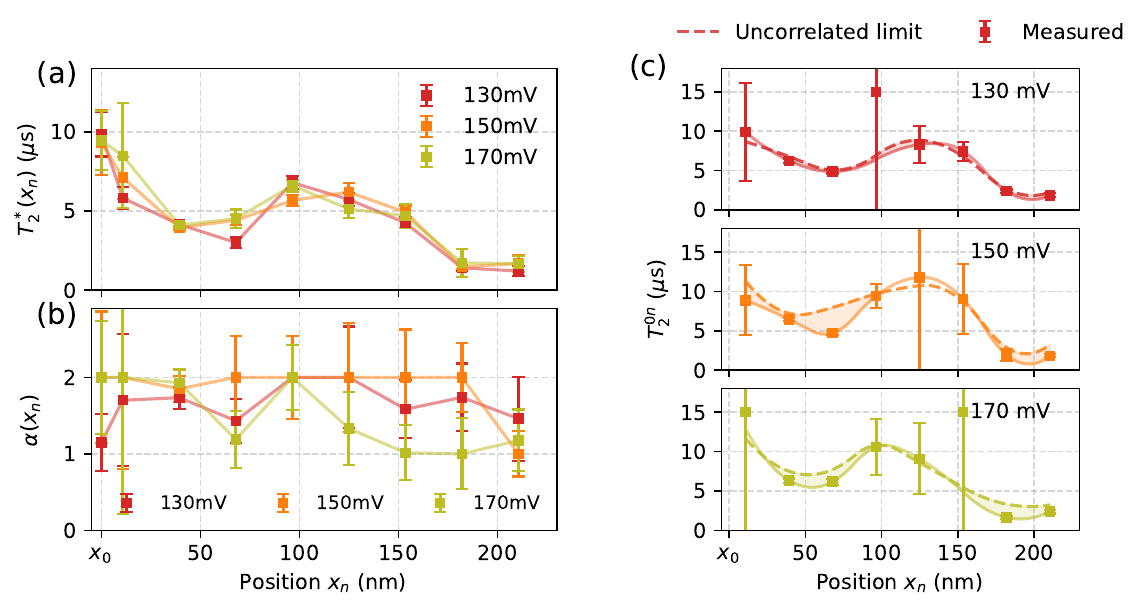}
    \caption{Consistency of noise characteristics. (a) Stationary decoherence time $T_2^*(x_n)$ measured at three conveyor voltages (130, 150, and 170 mV). (b) Two-point Ramsey coherence times $T_2^{0n}(x_n)$ for the corresponding voltages. The measured data (squares) is compared to an uncorrelated noise model (dashed lines) derived from the stationary data in (a). Shaded regions indicate the magnitude of spatial noise correlations. }
    \label{fig:figS_consistency}
\end{figure}

In Fig.~\ref{fig:figS_consistency}~(a) we show three different datasets taken in the high-field regime. With different colors we show decay times $T_2^*(x_n)$ taken with different conveyor voltages: 130 mV, 150 mV, and 170 mV, with the middle value corresponding to the two-point measurements in Figs~\ref{fig2:noise_features} and \ref{fig3_twopoint}~(a) and (c) of the main text. 

In Panel (a), we observe a consistent trend of the stationary $T_2^*(x_n)$ across all voltages. Similarly, also the two-point Ramsey dephasing times shown in Panel (c) consistently show  the same main trend across the three measurements, despite small differences in individual data points and some outliers. However, there is noticeable variability in the temporal correlations associated with the decay, as seen from the extracted exponents $\alpha(x_n)$ in Panel (b). The $150$~mV data (orange) consistently shows an exponent of $\alpha(x_n) \approx 2$, pointing at a dominant role of low-frequency noise everywhere, apart from the last point above $200$~nm. For the strongest confinement ($170$~mV), the decay above $x_n = 100$~nm becomes characterized by $\alpha(x_n) \to 1$, indicating a dominant contribution from high-frequency noise. Interestingly, the temporal correlations are also reduced for the weaker confinement for which the exponent remains significantly below $2$ for most of the sampled points.

\newpage

\section{Stationary and Two-point experiment data}
\label{app:two_points_data}

In this appendix we present the supplementary data of the stationary and two-point experiment, used in Figs.~\ref{fig2:noise_features} and \ref{fig3_twopoint}. 
\begin{figure}[htb!]
    \centering
    \includegraphics[width=0.90\linewidth]{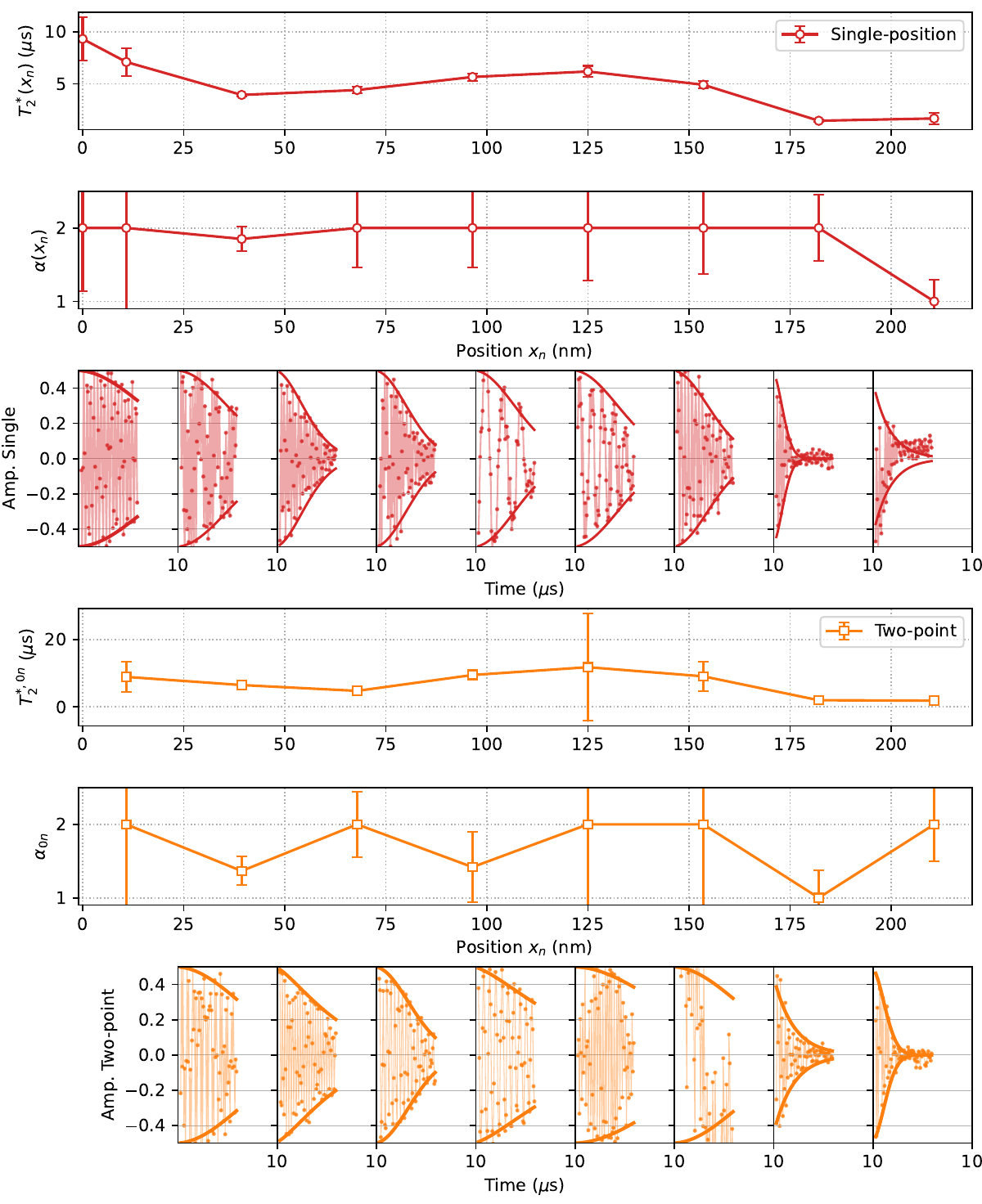}
    \caption{Two point experiment in high-field regime. With different colors we show coherence decay at single position $T_2^*(x_n)$ (red), and two-point decoherence $T_2^{AB}$ (orange). For each of them first row depicts corresponding decoherence time, second decoherence exponent and in the third one each column correspond to measured return probabilities at each point, with fitted envelope function.}
    \label{fig:placeholder}
\end{figure}

\begin{figure}[htb!]
    \centering
    \includegraphics[width=0.90\linewidth]{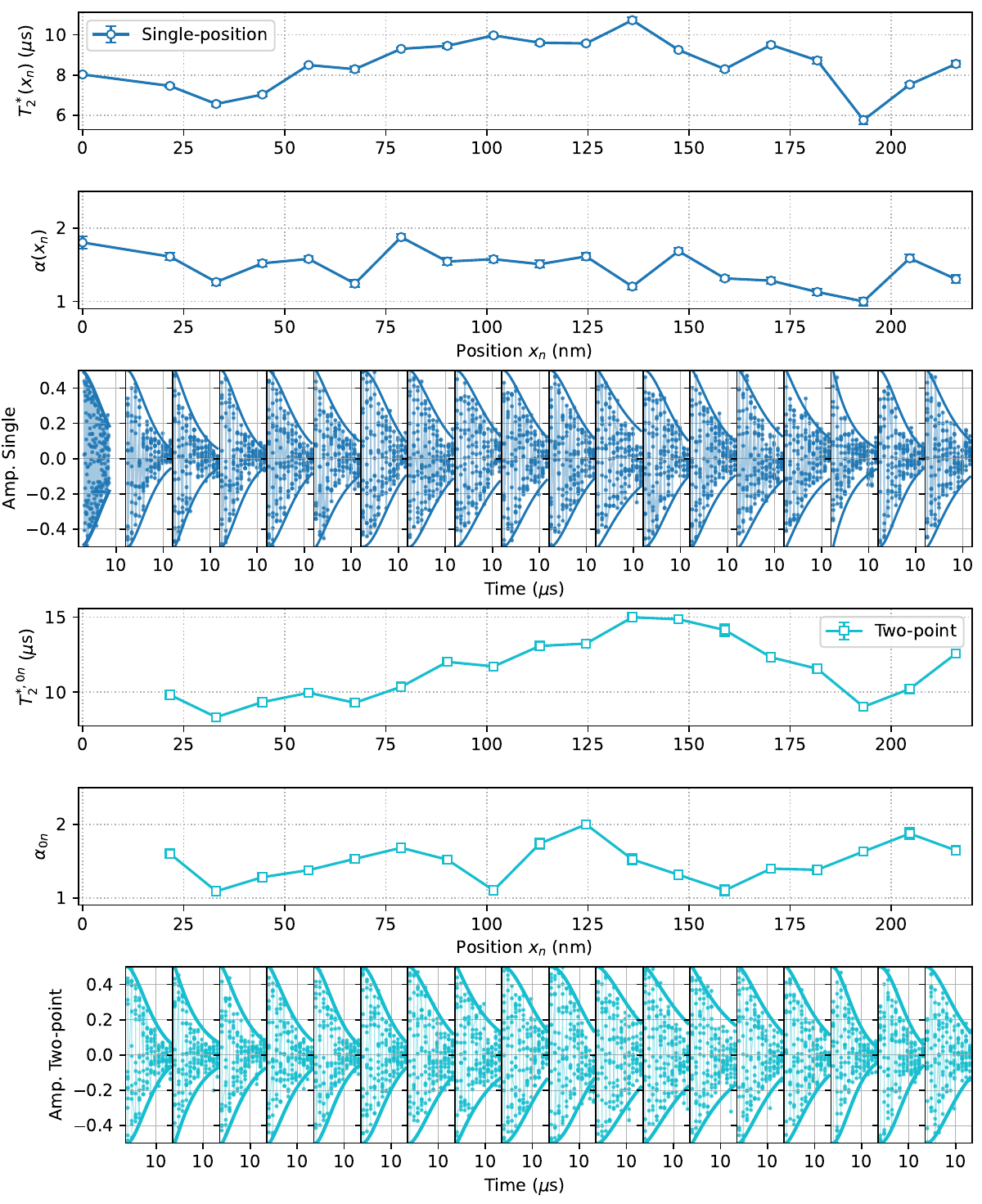}
    \caption{Two point experiment in low-field regime With different colors we show coherence decay at single position $T_2^*(x_n)$ (blue), and two-point decoherence $T_2^{AB}$ (cyan). For each of them first row depicts corresponding decoherence time, second decoherence exponent and in the third one each column correspond to measured return probabilities at each point, with fitted envelope function.}
    \label{fig:placeholder}
\end{figure}
\newpage \,
\newpage

\section{Periodic shuttling experiment data with and without DD pulses\label{app:shuttling_data}}
In this appendix we present the supplementary data of the shuttling experiment, related to Fig.~\ref{fig4_shuttling}. 
\begin{figure}[htb!]
    \centering
    \includegraphics[width=1\linewidth]{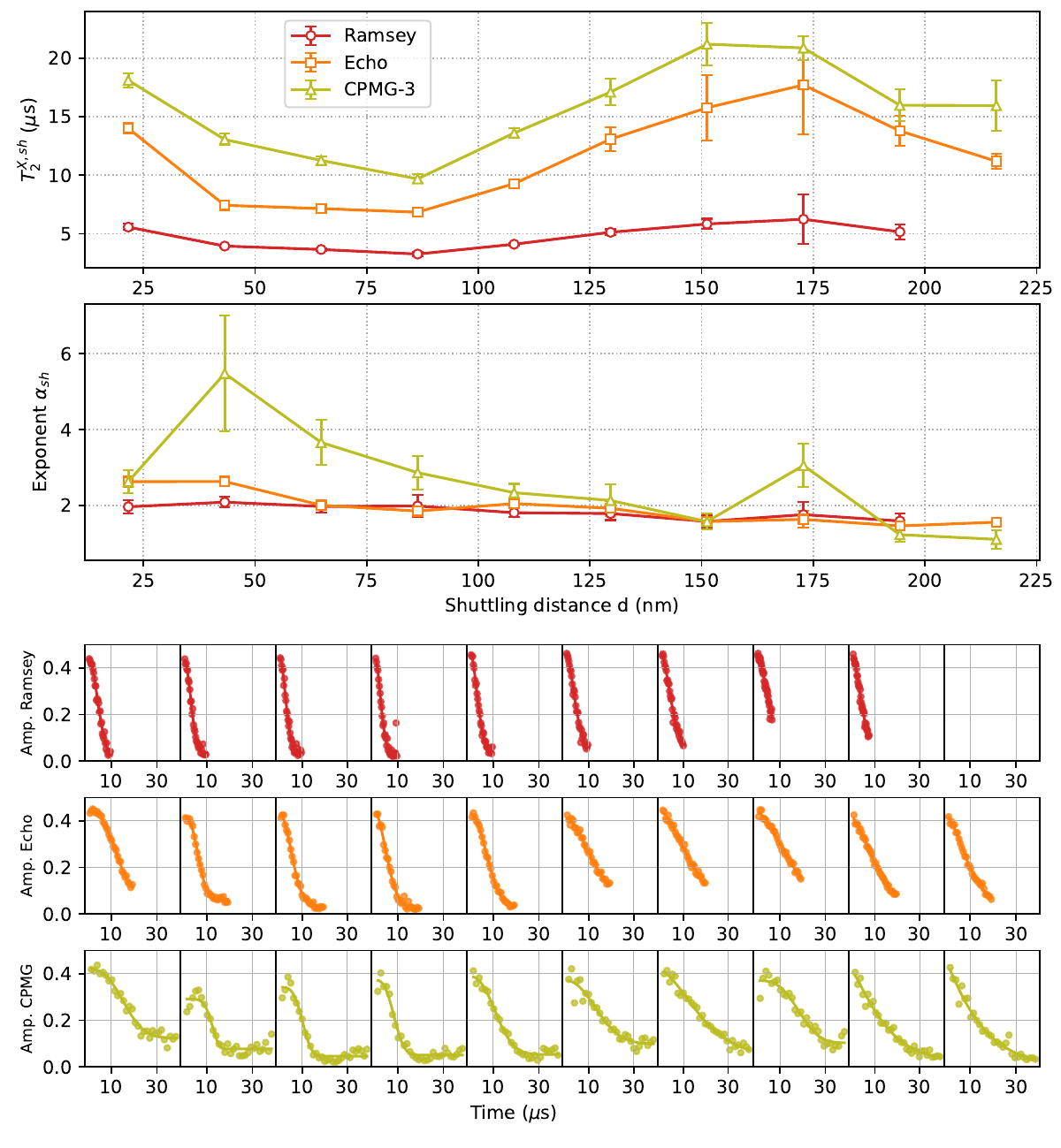}
    \caption{Shuttling Experiment at high-field. Using different colors we show Ramsey (red), Hahn-Echo (orange) and CMPG-3 (olive) experiments. First row depicts $T_{2}^{sh}$, the second $\alpha_{sh}$. The raw data and the fits are presented in the three bottom rows.}
    \label{fig:app_high_field_data}
\end{figure}
\begin{figure}[htb!]
    \centering
    \includegraphics[width=1\linewidth]{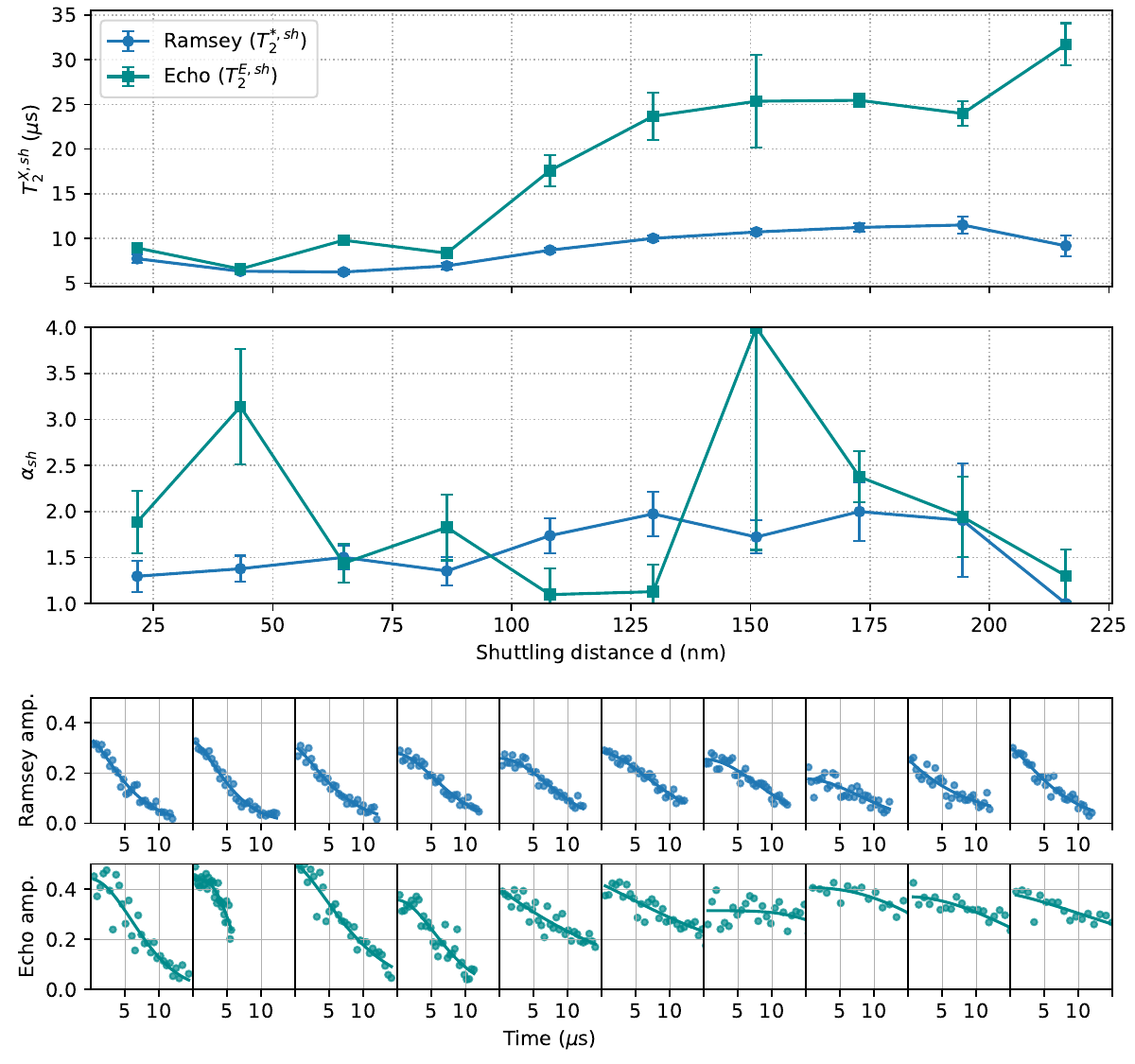}
     \caption{Shuttling Experiment at low-field.  Using different colors we show Ramsey (blue), Hahn-Echo (cyan) experiments. The first row depicts $T_{2}^{sh}$, the second $\alpha_{sh}$. The raw data and the fits are presented in the two bottom rows.}
    \label{fig:app_low_field_data}
\end{figure}

\newpage

\end{document}